\documentclass[journal]{IEEEtran}
\ifCLASSINFOpdf
\else
\fi
\usepackage{cite}
\usepackage{amsmath,amssymb,amsfonts}
\usepackage{algorithmic}
\usepackage{graphicx}
\usepackage{textcomp}
\usepackage{float}
\usepackage{pdfpages}
\usepackage{array}
\usepackage{tabu}

\usepackage{cite}
\usepackage{amsmath,amssymb,amsfonts}
\usepackage{algorithm,algorithmic}
\usepackage{graphicx}
\usepackage{xcolor}
\usepackage{algorithmic}
\usepackage{graphicx}
\usepackage{caption}
\usepackage{amsmath}
\usepackage{csquotes}
\usepackage{tabu}
\usepackage{lineno,hyperref}
\usepackage{amsthm}
\usepackage{mathtools}
\usepackage{subcaption}

\newtheorem*{remark}{Remark}

\DeclareMathAlphabet{\mathcal}{OMS}{cmsy}{m}{n}

\usepackage{cite}
\begin{document}
\title{
Drive Safe: Cognitive-Behavioral Mining for Intelligent Transportation Cyber-Physical System
}

\author{Md.~Shirajum~Munir,~\IEEEmembership{Student~Member,~IEEE,}
	Sarder~Fakhrul~Abedin,~\IEEEmembership{Student~Member,~IEEE,}
	Ki~Tae~Kim,
	Do~Hyeon~Kim,
	Md.~Golam~Rabiul~Alam,~\IEEEmembership{Member,~IEEE,}
	and~Choong~Seon~Hong,~\IEEEmembership{Senior~Member,~IEEE}% <-this % stops a space
%\thanks{A preliminary version of this work \cite{IEEEhowto:Munir_GC_Multi_Agent} was submitted to IEEE Globecom 2019 and has been accepted for presenting in session "SAC SGC3: Smart Energy Management", December 11, 2019.}
\thanks{Md. Shirajum Munir, Ki~Tae~Kim, Do~Hyeon~Kim, and Choong Seon Hong are with the Department of Computer Science and Engineering, Kyung Hee University, Yongin-si 17104, Republic of Korea (e-mail:munir@khu.ac.kr; glideslope@khu.ac.kr; doma@khu.ac.kr; cshong@khu.ac.kr).}% <-this % stops a space <-this % stops a space
\thanks{Sarder Fakhrul Abedin is with the Department of Information Systems and Technology, Mid Sweden University, Sundsvall 851 70, Sweden, and also with the Department of Computer Science and Engineering, Kyung Hee University, Yongin-si 17104, Republic of Korea,(e-mail:sarder.abedin@miun.se).}
\thanks{Md.~Golam~Rabiul~Alam is with the Department of Computer Science and Engineering, BRAC University, Dhaka, Bangladesh, and also with the Department of Computer Science and Engineering, Kyung Hee University, Yongin-si 17104, Republic of Korea,(e-mail:rabiul.alam@bracu.ac.bd).}
\thanks{Corresponding author: Choong Seon Hong (e-mail: cshong@khu.ac.kr)}}

\maketitle

\begin{abstract}
This paper presents a cognitive behavioral-based driver mood repairment platform in intelligent transportation cyber-physical systems (IT-CPS) for road safety. In particular, we propose a driving safety platform for distracted drivers, namely \emph{drive safe}, in IT-CPS. The proposed platform recognizes the distracting activities of the drivers as well as their emotions for mood repair. Further, we develop a prototype of the proposed drive safe platform to establish proof-of-concept (PoC) for the road safety in IT-CPS. In the developed driving safety platform, we employ five AI and statistical-based models to infer a vehicle driver's cognitive-behavioral mining to ensure safe driving during the drive. Especially, capsule network (CN), maximum likelihood (ML), convolutional neural network (CNN), Apriori algorithm, and Bayesian network (BN) are deployed for driver activity recognition, environmental feature extraction, mood recognition, sequential pattern mining, and content recommendation for affective mood repairment of the driver, respectively. Besides, we develop a communication module to interact with the systems in IT-CPS asynchronously. Thus, the developed drive safe PoC can guide the vehicle drivers when they are distracted from driving due to the cognitive-behavioral factors. Finally, we have performed a qualitative evaluation to measure the usability and effectiveness of the developed drive safe platform. We observe that the P-value is $0.0041$ (i.e., $< 0.05$) in the ANOVA test. Moreover, the confidence interval analysis also shows significant gains in prevalence value which is around $0.93$ for a $95\%$ confidence level. The aforementioned statistical results indicate high reliability in terms of driver's safety and mental state.
\end{abstract}


% \fi
\IEEEpeerreviewmaketitle

\section{Introduction}
\label{sec:introduction}
\IEEEPARstart{A}{ccording} to the global status report on road safety by the World Health Organization (WHO), around 1.35 million people die each year due to road traffic crashes \cite{world2018global}. 
Therefore, the 2030 Agenda for \emph{Sustainable Development} has set an unprecedented goal of reducing the global number of injuries from road traffic crashes by half within 2020.
In fact, among several key factors for road traffic crashes, the issue of a distracted driving behavior \cite{fountas2019factors} is more prevalent with the growing concern over driver activity and psychological state.
For example, the WHO global status report on road safety indicates the use of the mobile phone during driving is approximately four times more likely to be involved in a crash than drivers not using a mobile phone.
Meanwhile, the use of Hands-free phones and texting while driving also impede ensuring the drivers' safety \cite{qiao2016state}.
Besides, the psychological states of the drivers affect the incidence of road collisions significantly \cite{alavi2017personality}.
As a result, the traditional transportation system has gone through rapid development which leads to the extensive research on intelligent transportation system \cite{munir2019towards}.

The cyber-physical systems (CPS) are the physical and engineered systems that enable the integration of the cyber-world of computing and communications with the physical world \cite{rajkumar2010cyber}.
As a result, the concept of intelligent transportation cyber-physical systems (IT-CPS) \cite{chang2020intelligent} has emerged from the traditional CPS, where the objective is to bridge the gap between the traffic information system and the physical transportation system in-terms of intelligent and efficient monitoring, coordination and control \cite{feng2017application}.
More specifically, the perceived physical transportation environment facilitates a deep understanding of the urban traffic for improving the current traffic conditions through the newly emerged computing, communication, and control technologies.  

%\textcolor{blue}{TODO: One paragraph}

Providing a safe driving environment in IT-CPS, it is imperative to make a bridge between cognitive-behavioral mining of the vehicle driver and emerged with the computing, communication, and control technologies of next-generation cyber-physical systems. Therefore, this study aims to identify the psychological factors so that drive safe platform can guide the vehicle driver when they are distracted from driving due to the cognitive-behavioral factors. In this regard, we propose an autonomous interaction between a vehicle driver's mental health and affective mood repairment during driving. To the best for our knowledge, researches are conducted individually for the vehicle driver's mental health recognition, and distracted behavior detection. In fact, affective mood repairment for the vehicle drivers is entirely unexplored. 

\begin{figure}[t]
	\centering
	\includegraphics[width=7.5cm,height=6.0cm]{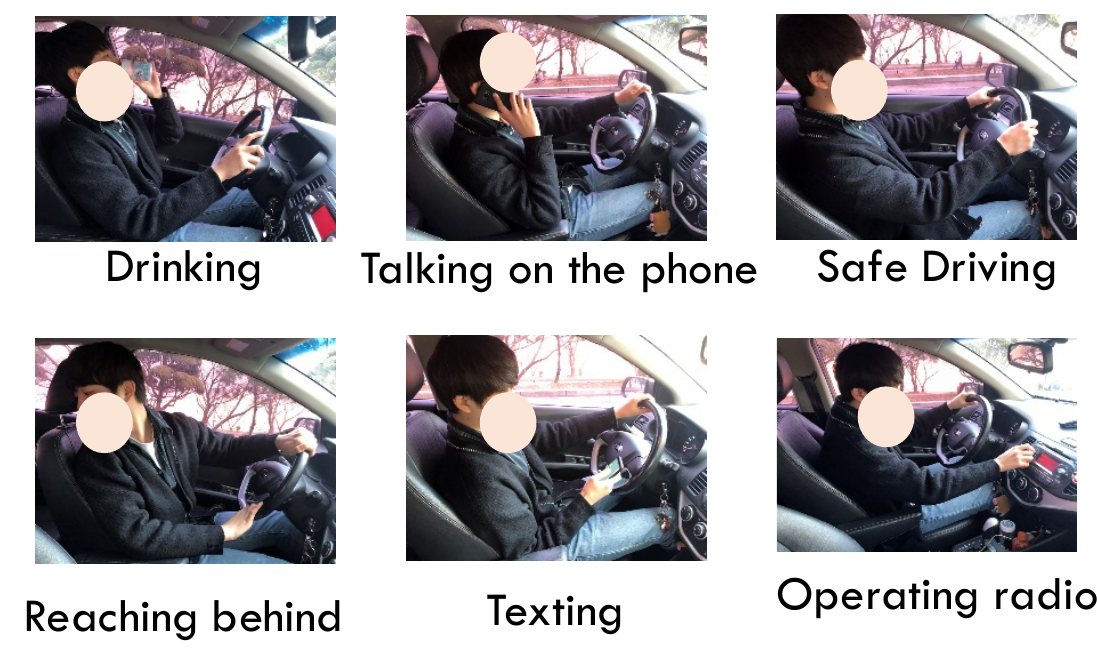}
	\caption{Example of driver activities during the drive \cite{munir2019towards}.}
	\label{Driver_Activity}
\end{figure} 

The main contribution of this paper is a novel drive safe platform that focuses on cognitive-behavioral mining of vehicle driver during the driving, promising to ensure the driver safety in intelligent transportation cyber-physical systems. Our key contributions include:

\begin{itemize}
	\item First, we propose the drive safe platform that is comprised of two established domains, cognitive engineering, and communication network for intelligent transportation cyber-physical system.   
	\item Second, we propose several artificial intelligence and statistical methodologies for the driver activity recognition, mood mining, cognitive-behavioral mining, and affective mood repairment of the vehicle driver. To repair affective mood, the system autonomously recommends a content (i.e., audio) based on the driver's current mental health. Therefore, the proposed models fulfill the vision of computing, communication, and control technologies by utilizing multi-access edge computing, on-devices computing, and autonomous decision making through the communication network.     
	\item Third, we develop a prototype of the drive safe platform for the intelligent transportation cyber-physical system that reduces the gap between the vision of the researcher community and industry. The developed prototype establishes the proof of concept (PoC) of the proposed drive safe platform.  
	\item Finally, a qualitative evaluation has performed to evaluate the usability and effectiveness of the developed drive safe platform. The statistical analysis has established the effectiveness of the developed drive safe platform. In particular, an average outcome is achieved more than $97\%$ for an upper $95\%$ confidence level (CL) of confidence interval analysis from the user studies. To this end, we examine the technical challenges and driven technologies for the drive safe platform toward road safety. We have found that the proposed cognitive-behavioral based road safety platform ensures the usability of the cutting-edge technologies for the future intelligent transportation cyber-physical systems.    
\end{itemize}

The rest of the paper is organized as follows. Section \ref{sec:Related_Works} presents related works of the drive safe platform in IT-CPS. The proposed drive safe platform in IT-CPS is described in Section \ref{sec:drive_safe_CPS}. We provide testbed implementation of the cognitive-behavioral mining for the drive safe platform in IT-CPS in Section \ref{sec:testbed}. Discussion and key findings are given in Section \ref{sec:discussion}. Finally, conclusions are drawn in Section \ref{sec:conclusion}.
\section{Related Work}
\label{sec:Related_Works}
\begin{figure}[!t]
	\centering
	\includegraphics[width=8.0cm,height=6.5cm]{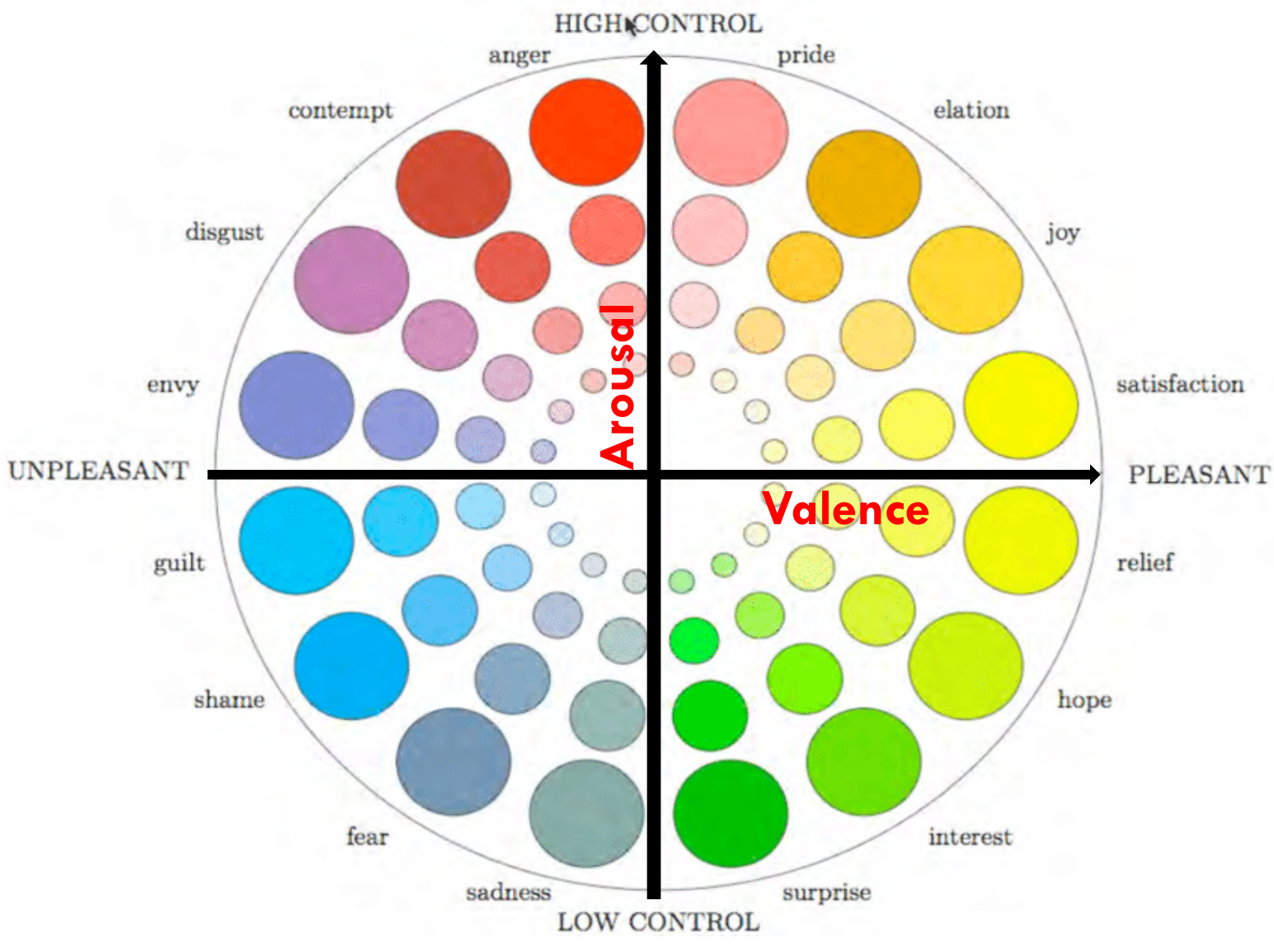}
	\caption{Geneva emotion wheel \cite{scherer2005emotions}.}
	\label{Geneva_Emotion_Wheel}
\end{figure} 
The intelligent transportation system (ITS) \cite{ferdowsi2019deep, khan2020edge, cheng20175g,chen2017vehicle, moioli2020neurosciences, liu2020contour, ndikumana2020deep, zhang2011data} is a prominent outcome of the next-generation of communication systems. To ensure the driver safety \cite{sadeghi2016safedrive, xing2019personalized, malinverno2020edge, ingale2020understanding} in the intelligent transportation system, drive activity \cite{gjoreski2020machine, liang2007real, arefin2019aggregating, xing2019driver, li2019driving}, and mental health  \cite{nvemcova2020multimodal, healey2005detecting, le2019evaluating, deshmukh2019characterization,  ali2016cnn, dehzangi2019eeg} play a crucial role. In this section, we discuss background of the intelligent transportation system, some of the related works, and challenges, which are grouped into three categories: (i) intelligent transportation system and driver safety, (ii) driver activity recognition, and (iii) driver mood recognition.        

\subsection{Intelligent Transportation System and Driver Safety}
The intelligent transportation system is an integrated technology aimed at delivering innovative services relating to various modes of transportation and traffic management. 
%Such technology allowing the consumers to be better intelligence and reliable transport network in safer, and more organized. 
Such technology is allowing the consumers to intelligence and reliable transport network in a safer, and organized manner.
Moreover, a bundle of emerging technologies \cite{zhang2011data} converges the goal of ITS. In particular, edge computing \cite{khan2020edge, munir2018intelligent}, edge analytics \cite{ferdowsi2019deep}, reliable communication that includes $5$G \cite{cheng20175g, chen2017vehicle} and beyond \cite{moioli2020neurosciences}. In fact, the emerging application such as autonomous and connected vehicle control \cite{xing2019personalized}, infotainment \cite{ndikumana2020deep}, autonomous road map management for emergency vehicles \cite{liu2020contour} are established the success of ITS. However, the driver safety in ITS is one of the essences to ensure the road safety as well as the vehicle driver psychological health. 

Recently, some of the challenges of driver safety have been studied in \cite{sadeghi2016safedrive, xing2019personalized, malinverno2020edge, ingale2020understanding}. In \cite{sadeghi2016safedrive}, the authors developed an autonomous transportation application that estimates the mental fatigue of a driver using an electroencephalogram (EEG) measurement to ensure the safe drive. This work has been predicted a collision probability to avoid the road accident by fusing driver mental state with the car parameters. The authors in \cite{xing2019personalized} proposed a personalized vehicle trajectory prediction mechanism to find the leading vehicle trajectory by analyzing the driving behaviors (i.e., aggressive, moderate, and conservative) to assures the road safety for the connected and autonomous vehicle. In \cite{malinverno2020edge}, the authors studied service based enhanced collision avoidance mechanism in the cellular network and proposed an edge-based framework for the road safety. The authors in \cite{ingale2020understanding} conducted
a questionnaire based survey of $770$ drivers and mine the behaviour of those vehicle driver based on standard statistical tests for developing a proactive road safety strategy. This statistical test includes driver behaviour in the traffic signal, dilemma zone analysis, and driver comprehension, where result showed income level, education level, driving experience, age, gender, and frequency are the key factors of a driver behaviour. However, all of these existing works \cite{sadeghi2016safedrive, xing2019personalized, malinverno2020edge, ingale2020understanding} moderately enhance the road safety. Apart from that, our proposed \emph{drive safe} platform provides a complete cyber-physical system for the intelligent transportation system that not only ensures the road safety but also repair the affective mood of the vehicle driver during the drive for the long term driving.       

\subsection{Driver Activity Recognition}
In recent years, distracted driving is one of the major causes of the road accident. Therefore, it is imperative to detect driver activity during the drive to achieve the goal of the road safety in IT-CPS. However, some of the individual research works \cite{gjoreski2020machine, liang2007real, arefin2019aggregating, xing2019driver, li2019driving} have been conducted to determine the driver activity without any connection with IT-CPS. In \cite{gjoreski2020machine}, the authors proposed an end-to-end deep learning (DL) model to detect the driver distraction by analyzing Electrodermal Activity (EDA) signal from the nasal and palm along with heart rate and breathing rate. The authors classified into two classes: 1) with distractions or 2) without distractions, and monitor the driver distraction in real-time. In \cite{liang2007real}, the authors developed an in-vehicle driver distraction detection system by developing support vector machines (SVMs) that used the eye movements and driving log. The authors in \cite{arefin2019aggregating} proposed a mixed convolutional neural networks (CNN) and Histogram of Oriented Gradients (HOG) featured based method to detect real-time driver distraction. In \cite{xing2019driver}, the authors developed a CNN-based neural network model that can classify seven types of driver activities into two classes. To increase the driver comfort and safety during the drive, the authors in \cite{li2019driving} classified the driving styles into three categories (i.e., low-risk, moderate-risk, and high-risk) from the operational images. Further, the authors analyzed the performance of the driving style classification through the deep convolutional neural network, long short-term memory (LSTM), and pre-trained-LSTM. However, these works \cite{gjoreski2020machine, liang2007real, arefin2019aggregating, xing2019driver, li2019driving} do not investigate the problem of distracted driving in IT-CPS, nor they account for feedback or safety instruction to the drivers. In this work, we provide real-time driver distraction detection and safety instruction to the driver by on-vehicle computing with IT-CPS. Further, we fuse the driver's activity with the cognitive behaviour for the betterment of the driver mental health during the drive.

\subsection{Driver Mood Recognition}
To ensure the road safety in ITS, driver stress and fatigue \cite{nvemcova2020multimodal} recognition is one of the fundamental challenges. However, very few researches \cite{healey2005detecting, le2019evaluating, deshmukh2019characterization, ali2016cnn, dehzangi2019eeg} focus on light to analyzing physiological data for the driver mood recognition. In \cite{healey2005detecting}, the authors first proposed a stress detection mechanism for the driver from the physiological sensor data. Additionally, they have presented a method to collect Electrocardiogram (ECG), Electromyogram (EMG), skin conductance, and respiration sensor date during the drive from the driver. The authors in \cite{le2019evaluating} evaluated a cognitive distraction in the simulation environment from the driver's eye tracking and tested in a driving  car with a $15$ km/h speed. In \cite{deshmukh2019characterization}, the authors considered the ECG physiological signal to characterize driver mental state with distracted driving. Further, the authors in \cite{ali2016cnn} captured the EDA, Skin Temperature, and ECG signal from the driver to differentiate the mood into four basic categories using CNN. Additionally, the authors proposed a Dempster-Shafer-based evidence theory to obtain a more robust emotional state for the driver. To monitor the driver attention during the drive, the authors in \cite{dehzangi2019eeg} analyzed electroencephalogram (EEG) signal and identified the impact of the key brain regions. These studies \cite{healey2005detecting, le2019evaluating, deshmukh2019characterization, ali2016cnn, dehzangi2019eeg} focused on detecting stress and strike to a limited scope that does not answer the question of \emph{"what is next?"}. Therefore, in this paper, we provide a cognitive behavioural mining platform for IT-CPS that not only recognizes the driver mood but also repairs affective mood toward the road safety.            

\section{Drive Safe Platform for Intelligent Transportation Cyber-Physical System}
\label{sec:drive_safe_CPS}
\begin{figure*}[!h]
%	\centering{\includegraphics[width=\textwidth,height=10.5cm]{fig/Platform.pdf}}
	\centering{\includegraphics[width=\textwidth]{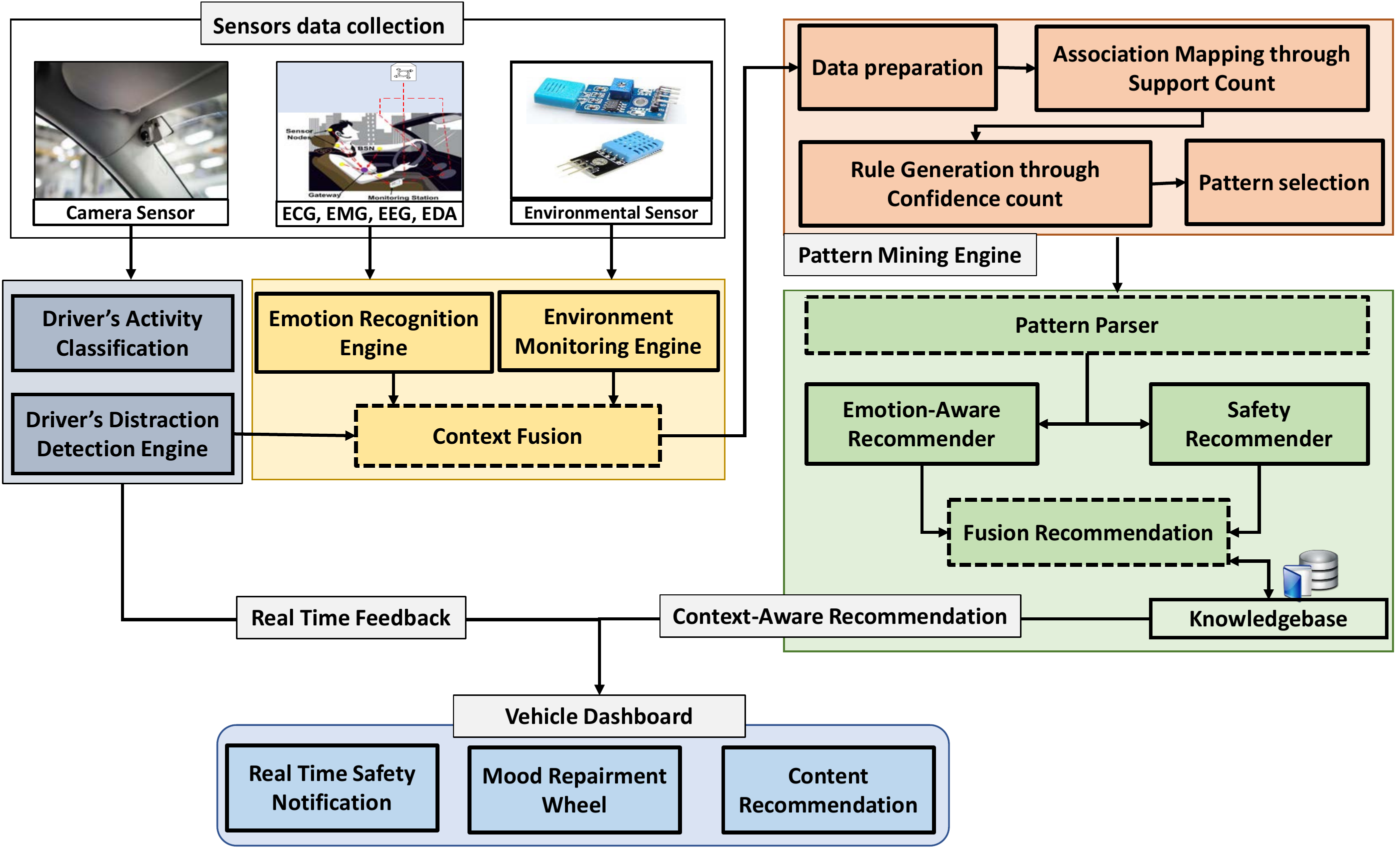}}
	\caption{Proposed drive safe platform for intelligent transportation cyber-physical system.}
	\label{Platform}
\end{figure*} 
The goal of the drive safe platform for intelligent transportation cyber-physical systems are to design and develop a driver safety application infrastructure with consideration of the driver's mental state while driving to prevent deadly road accidents, and injuries. In particular, we develop a personalized smart transportation safety platform using the user's biosensors, on-vehicle, and environmental sensors data as user lifelog. Thus, we employ cognitive-behavioral mining of the driver mood during the driving where mental health assessment and content recommendation based on vehicle environment and driver's emotion are considered for the mood repairment of the vehicle driver.

\subsection{Drive Safe Platform}
As we can see in Fig \ref{Platform}, the drive safe platform consists of several components, such components including various sensors, techniques, and methodologies. In the drive safe platform, we consider three sensors, such as biosensors (i.e., physiological and non invasive), on-vehicle camera sensors, and  environmental sensors. Thus, the collected biosensor data will be fed into the emotion recognition engine for the vehicle driver's mood mining. The emotion recognition module will use the CNN \cite{alam2019healthcare} for mood classification. Further, the in-vehicle camera and environmental sensor data are fed into a driver's activity classification and environment monitoring engine. The real-time camera sensor data is used for classifying the driver activity (e.g., safe driving, texting, talking on the phone, drinking, hair, and makeup, talking to passengers) during the driving via a driver activity classification module. Meanwhile, the driver distraction \cite{munir2019towards, xing2019driver} is classified into distracted or not distracted by the driver's distraction detection engine. Subsequently, the environment monitoring engine monitors the inside environment of a vehicle using environmental sensor data. The driver's emotional, activity, and environmental context are fused in the context fusion module and sent to the pattern mining engine.  

The pattern mining engine is deployed for association rule mining that finds the vehicle driver's lifestyle patterns. In this work, we design an association rule mining using an Apriori \cite{agrawal1993mining} method to extract cognitive patterns from observed sensor data. The support and confidence thresholds are used to find out the significant patterns. The pattern mining module sends a significant lifestyle pattern to the emotion-aware recommender. In particular, the pattern parser parses the context of the driver's lifestyle and sends it to the emotion aware recommender system for analyzing the emotional pattern by utilizing previous historical observation. The emotion-aware recommender system uses the Bayesian \cite{friedman1997bayesian} recommendation algorithm to recommend contents for mood repairment. 
% Meanwhile, a safety recommender is deployed based on Ripple Down Rules (RDR) \cite{compton1988knowledge} to recommend the activity for the vehicle driver’s physical and mental well-being. 
The fusion recommender fuses the driver's emotional with physical well-being recommendation and send it to the knowledgebase. Thus, a context-aware recommender of the smart on-board vehicle application enables the transportation safety notification and content to the vehicle driver. The smart on-board application keeps the driver's record on lifestyle patterns. That establishes the corresponding recommendation to the vehicle driver and provides an interface to interact through the smart transportation system. In particular, the vehicle dashboard provides feedback to the driver for the mood repairment and safe drive.

\subsection{Communication and computation infrastructure for the drive safe platform}
\begin{figure*}[!h]
	\centering
\includegraphics[width=\textwidth]{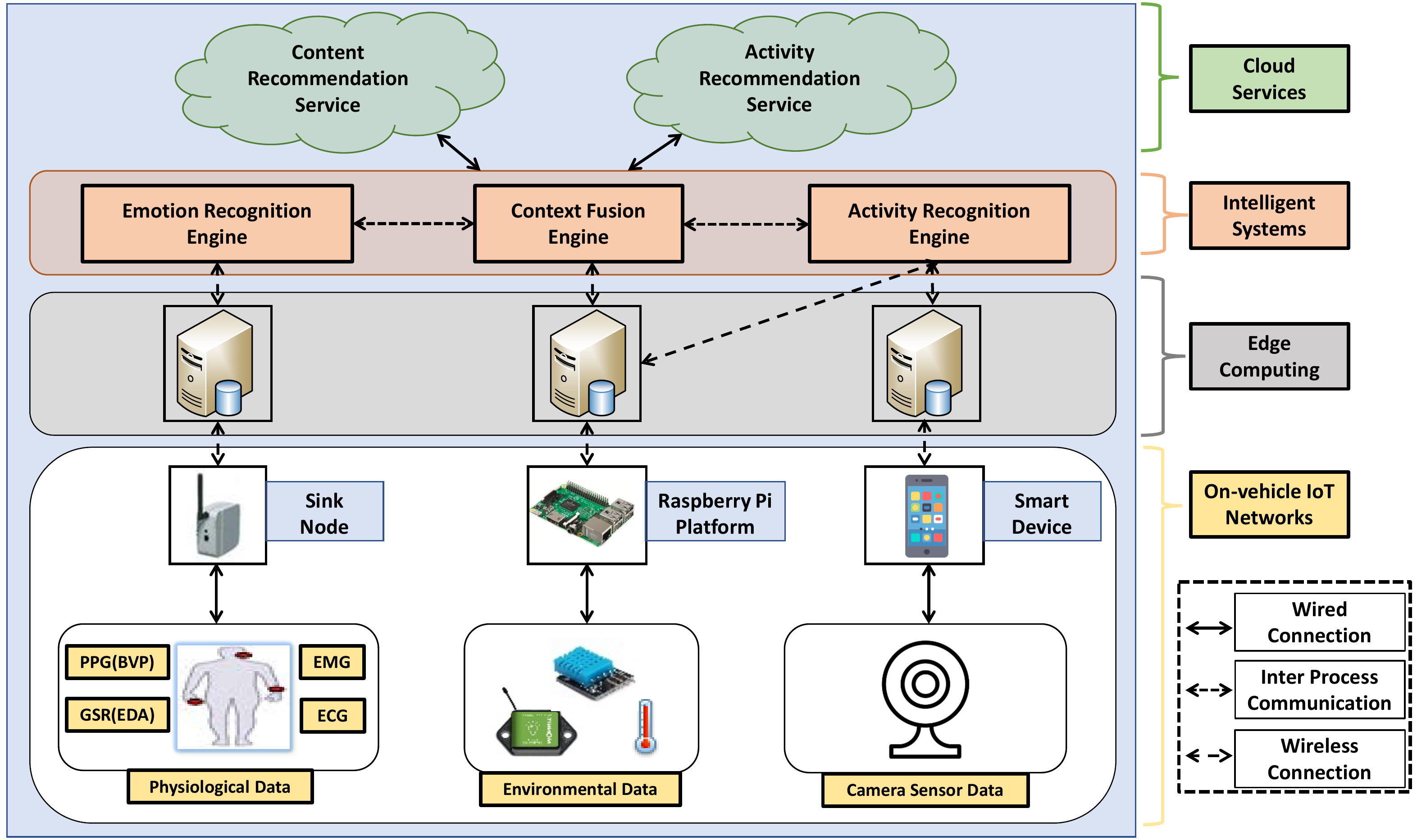}
	\caption{Communication and computation infrastructure for the drive safe platform in intelligent transportation cyber-physical system.}
	\label{communication_model}
\end{figure*} 

The communication and computation infrastructure (CCI) for the drive safe platform is considered as multi-access \cite{porambage2018survey, kekki2018mec, munir2019edge, alam2019edge} technology for the $5$G and beyond, as shown in Fig. \ref{communication_model}. In this communication model, we consider a four-layer communication and computation infrastructure. Such infrastructure includes on-vehicle IoT network, edge computing, intelligent system, and cloud services layer. In particular, this CCI can provide an intelligent transportation cyber-physical system by enabling on-board (i.e., on-device) computing, edge computing, and cloud computing.    

\subsubsection{On-vehicle IoT Network}
In this layer, we consider three vertical sensor networks with on-device computational capabilities. In particular, the on-vehicle IoT network \cite{abedin2015system} encompasses physiological, environmental, and inertial sensors networks.
\paragraph{Body Area Network}
In order to capture the physiological observation of a vehicle driver, we consider four types of biosensors sensors: electrodermal activity (EDA), electrocardiogram (ECG), electromyography (EMG) and Electroencephalography (EEG). These sensors capture the physiological \cite{alam2016psychiatry} data for the vehicle driver through a sink node. Besides, the sink node acts as a physiological data aggregator. Thus, the communication between sensors and the sink node (i.e., local and on-board) is Bluetooth IEEE Std 802.15.1 \cite{kinney2006ieee}. After the aggregation of the sensors data in the sink node, it communicates with the edge server over WiFi 802.11n-2009 standard \cite{ieee2007std}.   

\begin{figure*}[!h]
	\centering{\includegraphics[width=\textwidth]{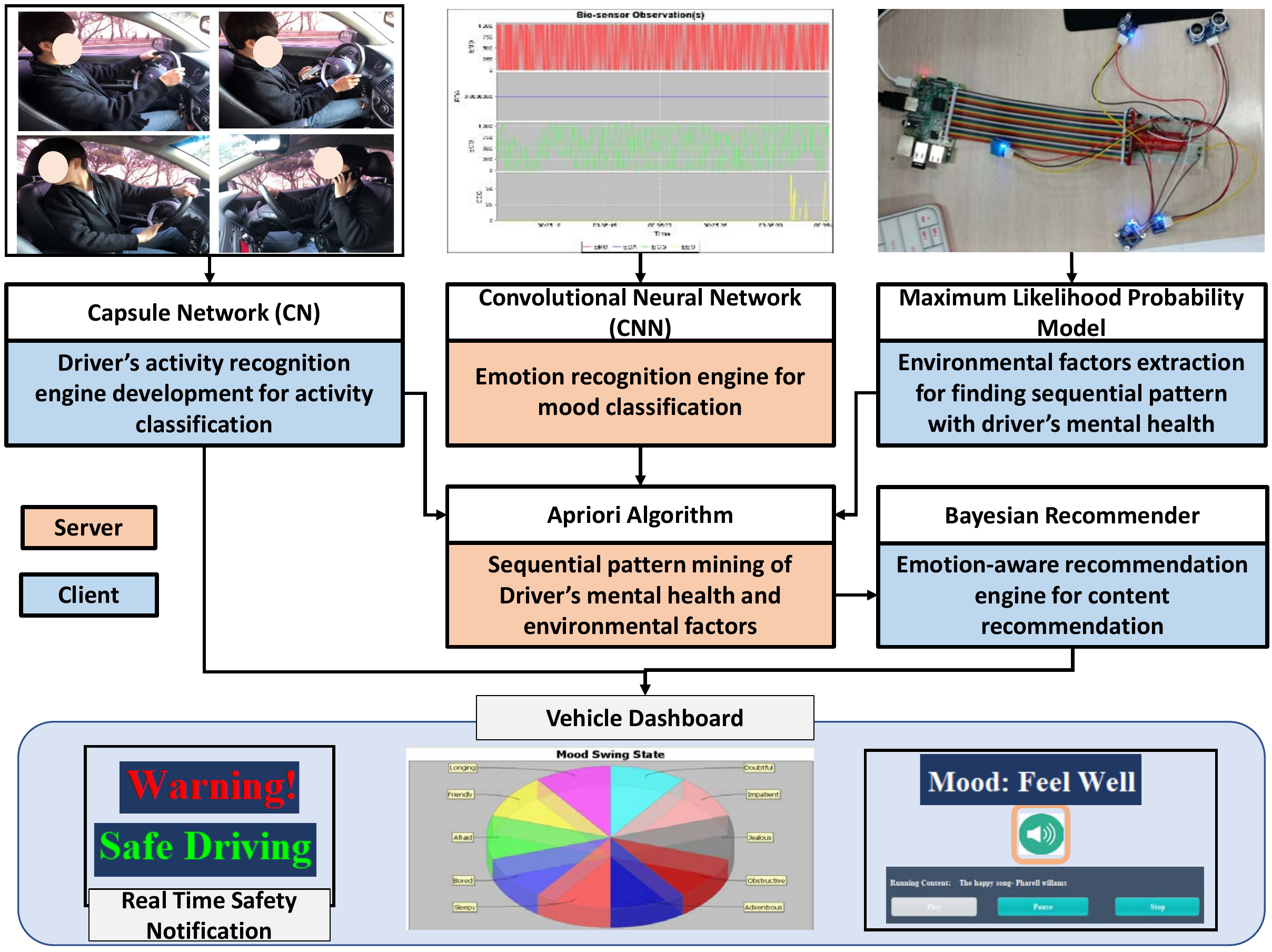}}
	\caption{Use case of cognitive behavior mining and recommendation system toward the drive safe in intelligent transportation cyber-physical system.}
	\label{Testbed_Implementation}
\end{figure*}

\paragraph{Environmental Sensors Network}
The environmental sensors network consists of three types of environmental sensors. We characterize the driver compartment's environment by capturing temperature, humidity, and ambient light sensor data. Thus, we consider a gateway node to collect those sensor data that can communicate to the edge server via a wireless LAN IEEE WiFi 802.11n-2009 standard \cite{ieee2007std}. This gateway node can preprocess and also extract features through the on-gateway computation.    

\paragraph{Camera Sensor Network}
The real-time video is captured by the camera sensor to detect the current activity of the vehicle driver while driving. This camera is physically deployed in the driver compartment and controlled by the on-board vehicle device. This on-board device can be any smart device that is attached to a car dashboard. Smart devices can communicate through the wireless LAN IEEE WiFi 802.11n-2009 standard to the multi-access edge server \cite{ieee2007std}.

\subsubsection{Multi-Access Edge Computing}
The multi-access edge computing (MEC) module is deployed in the second layer (bottom-up manner) of the drive safe platform in the IT-CPS. Therefore, this layer can communicate through multiple communication protocol for first computation, and low latency access \cite{taleb2017multi}. In the proposed communication and computation infrastructure, the MEC can communicate through Bluetooth, wireless LAN, and wired LAN for inter-layer communication. The communication protocols are not limited to the above protocols while MEC can communicate through any wireless communication \cite{abedin2018resource,bairagi2018qoe} such as LTE, 5G. Further, edge computing server provides the computation of several artificial intelligence \cite{munir2019artificial} methods that detect the vehicle driver's behavior and distraction for enabling the prevention mechanism. Thus, multi-access edge computing ensures heterogeneity in terms of both communication and computation, while several intelligent systems are decoupled from each other. 

%\subsubsection{Intelligent System}
\subsubsection{Intelligent Cyber-physical Transportation Systems}
To assure the driver safety through the intelligent cyber-physical transportation systems, it is imperative to accommodate the computation of driver emotion recognition, current activity recognition, and context-aware recommendation on the edge. Therefore, at the top of the multi-access edge computing, we consider an intelligent system layer that consists of emotion recognition, activity recognition, and context fusion engines. Each decision-making engine performs domain-specific AI tasks to ensure driver safety while distractions occur during the vehicle drive. These engines are deployed in a multi-access edge computing platform and the edge servers communicate with the cloud. In particular, a wired backhaul \cite{jaber20165g} connection between multi-access layer to the cloud.

\subsubsection{Cloud Services}
To facilitate a safe drive in the considered intelligent transportation system, we design a cloud service layer. In this work, this layer encompasses the content recommendation service and activity recommendation service. These services ensure the vehicle driver safety as well as employing driver's mood repairment for a safe drive. In particular, the integration of the cloud-based recommendation service ensures the robustness of the drive safe platform.

\subsection{Use Case of the Drive Safe Platform}
The cognitive behavior of a vehicle driver relies on mental health as well as physical behavior. Further, environmental factors also have a crucial role in mental health. Therefore, in this section, we illustrate the use of the drive safe platform that can reduce the chance of road accident by intelligently analyzing the vehicle driver's cognitive behavior. 

\begin{table}[!h]
	\caption{Driver activity class \cite{driverdataset}}
	\begin{center}
		\begin{tabular}{|c|c|}
			\hline
			\textbf{Description}&{\textbf{Class}} \\
			\hline
			Drive with concentration & Safe driving \\
			\hline
			Texting - right & Distracted driving\\
			\hline
			Talking on the phone - right & Distracted driving\\
			\hline
			Texting - left & Distracted driving\\
			\hline
			Talking on the phone - left & Distracted driving\\
			\hline
			Operating the radio & Distracted driving\\
			\hline
			Drinking & Distracted driving\\
			\hline
			Reaching behind & Distracted driving\\
			\hline
			Hair and makeup & Distracted driving\\
			\hline
			Talking to passenger & Distracted driving\\
			\hline
		\end{tabular}
		\label{tab:tab_driver_activity_class}
	\end{center}
\end{table}

Consider a scenario where a vehicle driver is driving a car with full of concentration without any mental stress. We assume that this vehicle driver is driving in a safe driving manner. However, during the drive, the driver can also check cell phone, talk to the other passenger, drink, operate the radio, or any other activities that induce distracted driving pattern (shown in Fig. \ref{Driver_Activity}). Therefore, we can classify the driver activity into ten categories as illustrated in Table \ref{tab:tab_driver_activity_class}. Besides, the driver activities also depend on the driver's current emotional state \cite{scherer2013grid}. In particular, according to Russell's emotion circumplex \cite{bradley1994measuring}, the human emotion relies on two vital factors: arousal, and valence. The arousal discretizes the driver's brain's physiological and psychological state that can stimulate a sense of organs to a print of perception. A higher value of arousal describes high anxiety and excitement, which is the intensity of the driver's physiological changes. Meanwhile, valence can characterize the driver's emotion into attractiveness and averseness. Notably, a positive value intuits goodness of emotion while negative describes the badness of the emotion. Thus, Fig \ref{Geneva_Emotion_Wheel} illustrates the relationship between arousal and valence with the human mood. In particular, the value of arousal and valence represents the driver’s mood and also relies on the indoor compartment environmental factors such as light intensity, temperature, and humidity. Therefore, to ensure drive safe in an ITS-CP system, the driver mood repairment is essential. To accomplish this goal, we design an use case that assures the driver safety during the driving. We illustrate the use case of cognitive behavior mining and recommendation system of the drive safe platform in Fig \ref{Testbed_Implementation}.       

In this use case, we consider server and client, where each vehicle acts as a client, and edge/cloud serves as a server in the intelligent transportation cyber-physical system. Particularly, a pre-trained driver's activity recognition \cite{munir2019towards}, maximum likelihood (ML) \cite{white1982maximum} models for on-compartment environment observation, and Bayesian \cite{friedman1997bayesian} content recommender for mood repairment are deployed in the on-vehicle device. Meanwhile, the edge server is classifying \cite{alam2019healthcare} the driver's emotion by physiological sensor observation of the driver. The on-vehicle device collects those observations through the physiological sensors that are placed in the driver’s body. Further, sequential pattern mining of driver current emotion and environmental factor will be performed by the server. Such a sequential pattern mining engine receives inputs as current activity, environmental context, and driver's emotion. The server then sends the sequential pattern mining result to the on-vehicle device for further processing. Thus, the on-vehicle device executes a Bayesian content recommender to select the content (i.e., audio songs) for the driver mood repairment. This content helps the vehicle driver for the mood repairment, reducing the driver's anxiety toward the safe drive. Subsequently, the on-vehicle device also notifies the safety message based on real-time driver activity recognition via the on-board vehicle display. The mental health of the particular vehicle driver improve based on recommended mood repairment contents. In the case of intelligent transportation cyber-physical system, each vehicle follows the same procedure in real-time. To this end, the Bayesian content recommender recommends personalized contents (i.e., audios) to each vehicle driver based on the personal choice that ensures the personalized behavior mining for the driver.

The cognitive behavior mining and recommendation system for the drive safe platform in IT-CPS, we have deployed five AI-based models. Capsule network (CN) \cite{sabour2017dynamic} is considered for the vehicle driver's activity recognition and classification. Meanwhile, a convolutional neural network recognizes human emotion \cite{alam2019healthcare}. Further, a statistical maximum likelihood (ML) \cite{white1982maximum} model determines the statistical environment factors for the on-compartment environment (i.e., light intensity, temperature, and humidity) of the vehicle. In order to conduct a behavior mining of the vehicle driver, we have deployed Apriori-based \cite{agrawal1993mining} sequential behavioral pattern mining model to capture the behavior of a driver by analyzing the driver mood, activity and environmental factor. Finally, a personalized Bayesian content recommendation \cite{friedman1997bayesian} engine is developed to repair a vehicle driver's mood. This ensures a safe driving environment in the IT-CPS. A detailed description of the testbed implementation is given in the later section.

\section{Testbed Implementation for Drive Safe Platform}
\label{sec:testbed}
\begin{figure*}[!h]
%	\centering{\includegraphics[width=\textwidth,height=10.5cm]{fig/Demo_testbed.pdf}}
	\centering{\includegraphics[width=\textwidth]{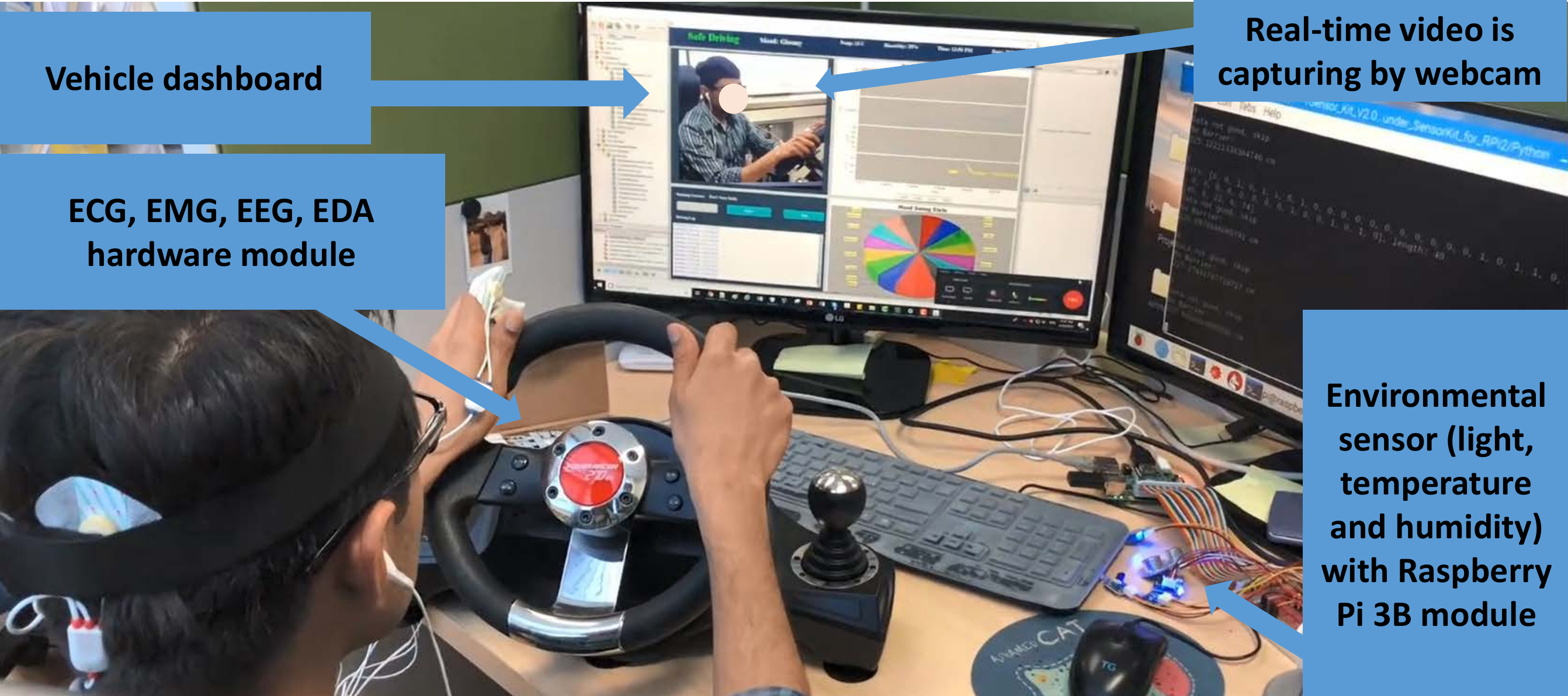}}
	\caption{Testbed for drive safe platform in intelligent transportation cyber-physical system.}
	\label{Demo_testbed}
\end{figure*} 
The drive safe testbed consists of several components, including various sensors with sink node, on-vehicle computational processing units, and edge server. In this work, we have deployed four kind of physiological sensors, three type of environmental sensors and a camera sensor to capture the physiological, environmental, and activity data, respectively. We consider Raspberry Pi 3B Model as a environment sensor gateway (Raspbian OS), a core i3-7100 processor with a speed of 3.9 GHz along with 8 GB of RAM as an on-vehicle device (Windows 10 OS), and  a core i7 processor with a speed of 2.6 GHz along with 32 GB of RAM as an edge server (Ubuntu 16.04 LTS OS). We illustrate the testbed environment in Fig \ref{Demo_testbed}.

\subsection{Sensor Data Collection and Preprocessing}
In this work, we collect observational data from eight sensors to determine the driver's mood. Meanwhile, we recommend the content to the vehicle driver for mood repairment that accomplishes the goal of drive a safe platform in IT-CPS toward driver safety. We have developed this data collection system by combining three IoT networks. A body area network (BAN) is created to collect physiological data of the vehicle driver. Further, we deployed environmental sensor data along with a gateway design. Finally, we capture real-time camera sensor data using an on-vehicle camera. A brief discussion of these sensor data collections and preprocessing mechanisms is discussed in the following section.  

\subsubsection{Physiological Sensors}
\begin{figure}[!h]
	\centering
	\includegraphics[width=8.7cm,height=5.5cm]{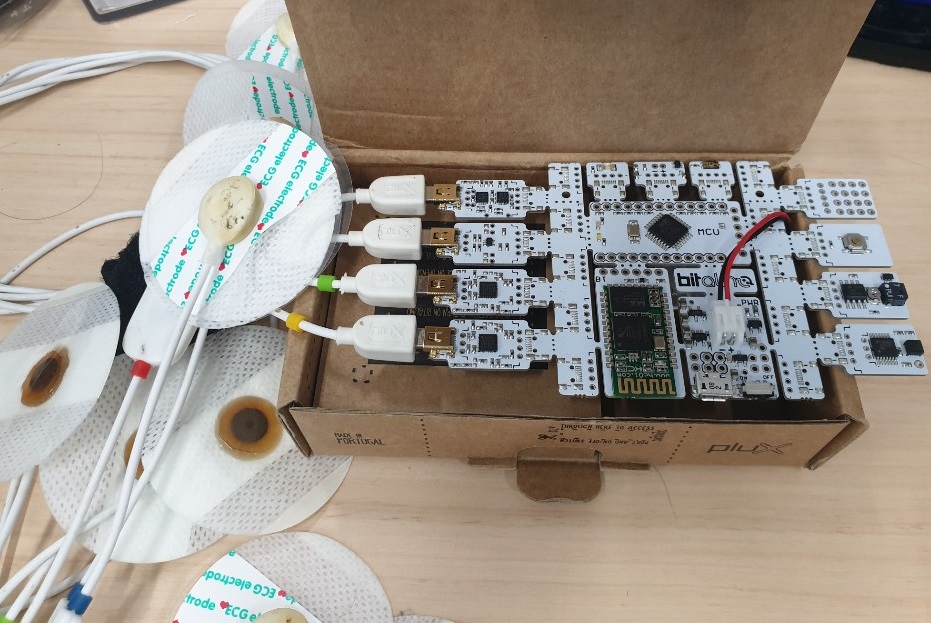}
	\caption{BITalino (r)evolution Board Kit BLE physiological sensors module \cite{Bio_sensor}.}
	\label{Bio_sensor_module}
\end{figure}
A BAN is formed to collect the physiological sensor data through a sink node. The physiological sensor communicate to the sink node via Bluetooth, and four sensors are attached with the wire connection to the sink node. Thus, to ensure the high accuracy measurements of the bio-information \cite{gravina2017multi, fortino2015framework} from the driver, we considered ElectroMyoGram, Electrocardiogram, Electro-dermal Activity, and Electroencephalography. The physiological sensors and sensor hub is depicted in Fig. \ref{Bio_sensor_module}. In particular, a BITalino (r)evolution Board Kit BLE \cite{Bio_sensor} physiological sensors module is used for BAN.

\paragraph{EMG}
The trapezius muscle's electrical activity is measured via the EMG \cite{vera2008wearable} surface, which comprises the prevalent characteristics of mental stress and irritation. As possible features of the EMG signal, the mean amplitude, mean and median frequency, average EMG gaps, and percentile of EMG gaps are extracted. At first, the signals received from the EMG are preprocessed for denoising and baseline removal and the applied sampling rate is $256$ Hz. The bandpass filtering approach is applied to cut off the low frequencies that removes the baseline wanders and also discards high-frequency noises. The upper trapezius muscles EMG is useful for measuring anxiety and stressful mode of the subject. Thus, the stress will appear when the sympathetic nervous system is activated. One of the consequences is elevating muscle tone, which sometimes leads to shivering. Capturing this muscle tone elevation could then be a predictor of the level of mental stress. During a stressful situation, the studied \cite{healey2005detecting, vera2008wearable} show significantly higher amplitudes of the EMG signals than rest and fewer gaps during stress. In particular, mean and median frequencies were significantly lower during stress than rest. EMG features correlated with subjectively indicated stress levels of the vehicle driver. 

The previous state-of-the-art study \cite{vera2008wearable} indicated induced stress, which resulted in changes in the trapezius muscle's EMG signals. These EMG changes included an increase of amplitude and a decrease in the number of recorded gaps. Both are indications of elevation of muscle activity, caused by the stress tasks \cite{wijsman2010trapezius}. However, by measuring the stress level, we can infer the subject's negative mood, which is an indication of negative valence and negative aerosol.

\paragraph{ECG}
To improve the accuracy of emotion classification, we collect ECG \cite{agrafioti2011ecg} sensor observation form the driver while driving and listening to audio stimuli. Therefore, we have analyzed the received ECG signals to extract discriminative features. To do this, we analyze the R-R intervals, QRS amplitude and duration, QT interval, QTc interval, R/S ratio, and R-peak from the driver's ECG signals. The mean, variance, first and second derivatives are also applied with respect to time. As a result, we can utilize those patterns as a functional feature strongly affects valence and aerosol.

We collect ECG signals through the sensor hub and then pre-processed for denoising and baseline removing. In particular, we have applied the sampling rate to $256$ Hz. The bandpass filtering approach is being used to cut off the low frequencies that eliminate the baseline wanders and discard high-frequency noises. Eventually, we do the smoothing by taking the $15$ sample window on the moving average.

\paragraph{EDA}
One of the most critical indicators for emotional arousal is the electrodermal function. The EDA is derived from the autonomic stimulation of skin sweat glands. The emotional pressure causes the sweating on hands and feet. The EDA data reveal distinctive patterns, which can be quantified statistically, while we are emotionally aroused. The obtained EDA bio-signals for the analysis of the emotions incorporate specific noise types. Using bandpass filtering, we filter the received signals from EDA biosensor that can isolate the noises. The electrical potential is applied on two points of skin contact to measure the Skin Conductance Level (SCL), and the current flow between those points is measured. The Skin Conductance Response (SCR) is calculated from two consecutive zero-crossing behaviors. The EDA is the most useful index that can identify emotional and cognitive states because parasympathetic behavior does not contaminate it.

\paragraph{EEG} 
To capture the brain activity for a particular event, we consider Electroencephalography \cite{alotaiby2015review} signal from the vehicle driver. We can find the voltage fluctuations occur by an ionic current within the driver’s brain's neurons.The EEG sensor's electrical signal discretizes a potential valence by placing it to the scalp of the driver. We capture this signal with a sampling rate of $8$ Hz.

\subsubsection{Environmental Sensor}
\begin{figure}[!h]
	\centering
	\includegraphics[width=8.7cm,height=5.5cm]{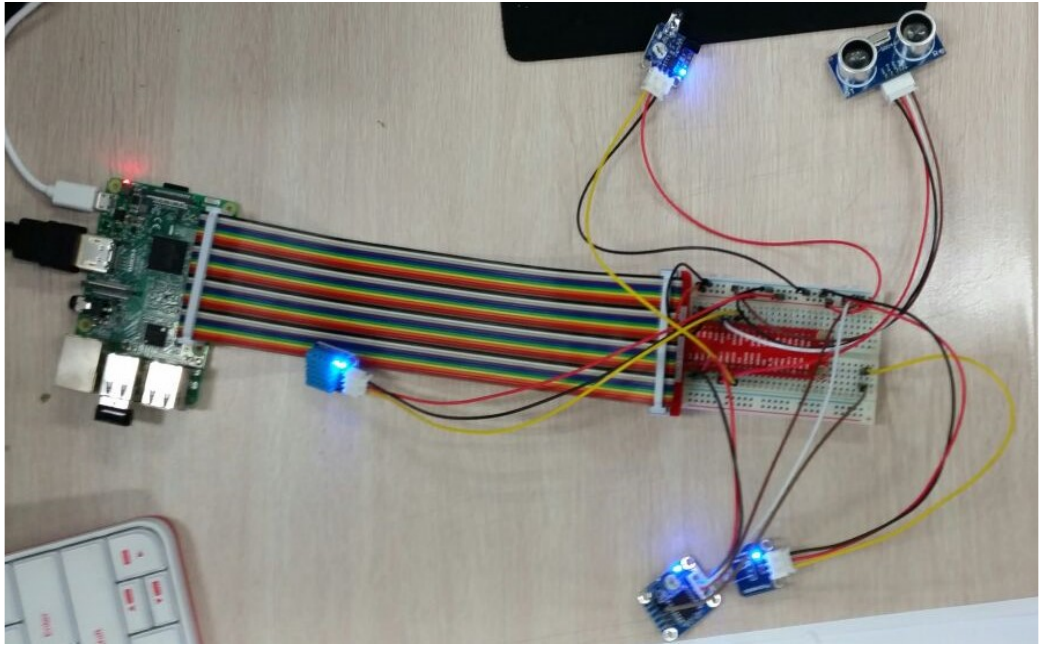}
	\caption{Testbed setup for environmental sensor with Raspberry Pi 3B module.}
	\label{pie_3_testbed}
\end{figure} 

\begin{figure}[!h]
	\centering
	\includegraphics[width=8.7cm,height=5.5cm]{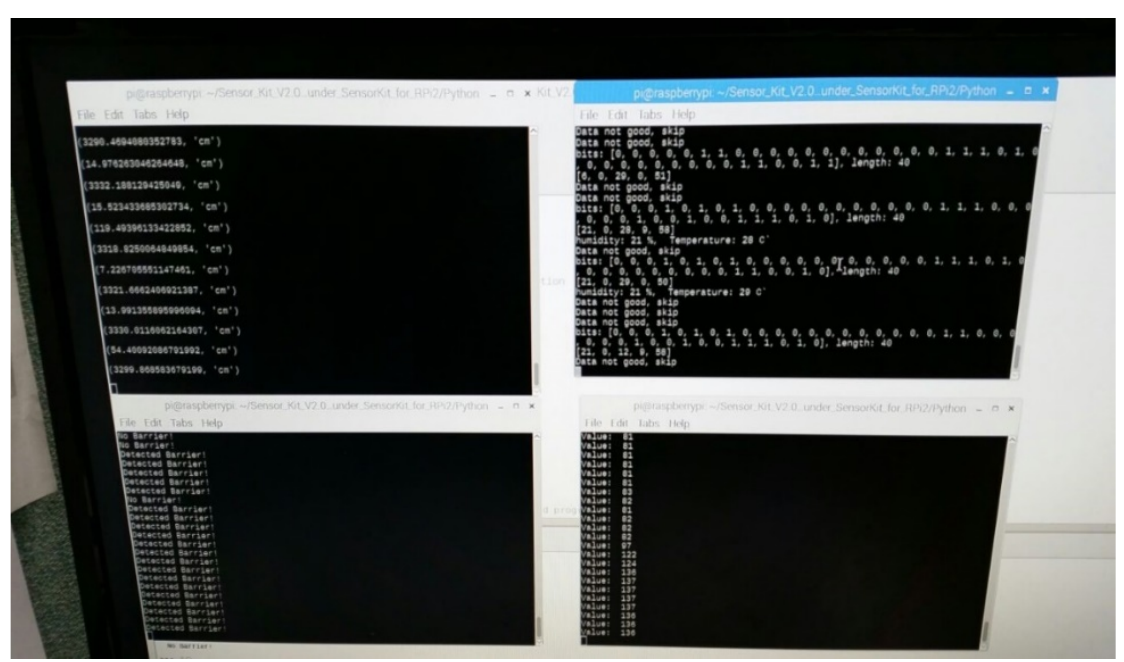}
	\caption{Environmental sensor data collection.}
	\label{environment_sensor_testbed}
\end{figure}
In the first sensor module, we describe how to collect physiological sensor data of the vehicle driver during the driving. To capture the on-compartment environment data, we consider an ambient light sensor, temperature, and humidity sensor. Thus, such sensors are attached with a Raspberry Pi 3B platform that acts as a gateway node of the considered cyber-physical system. In Figs \ref{pie_3_testbed} and \ref{environment_sensor_testbed}, we illustrate the environmental sensor data acquisition system. In this environmental data acquisition system, the sensory observations are preprocessed and the features are extracted from the collected environmental sensory data. In particular, the environmental sensors (e.g., ambient light sensor, temperature and humidity sensor) data are recorded and executes the maximum likelihood probability model for preprocessing. We capture $30$ data points for each sensor in one observational period.

\paragraph{Ambient Light Sensor}
The ambient light sensor is used to detect the brightness of the driver's compartment and thus, provides an ambient environment. The features are extracted for each time window, such features include Mean (average of the data in the time window), Min (Minimum value in the time window), Max (Maximum value in the time window), Std (Standard deviation of the data in the time window), End-Start diff (the difference between the last value and the first value in the time window), and Max-Min diff (the difference between the maximum and minimum values in the time window).

\paragraph{Temperature Sensor}
The temperature sensor can capture the current temperature of the driver's compartment environment. In particular, the extracted features of the sensor are time, sample number, temperature, and voltage.

\paragraph{Humidity Sensor}
The humidity sensor indicates the likelihood of precipitation, dew or fog in the user environment. The extracted features of the sensor are RH (Relative Humidity), PPM (Parts Per Million), D/F PT (Dew/Frost Point), AB (Absolute Humidity), and timestamp.

\subsubsection{Camera Sensor}
To capture the vehicle driver activity image, we consider webcam with $640 \times 480$ resolutions. We capture the image frame with an interval of $100$ ms. This image directly stores in the on-vehicle device and we preprocess as $128 \times 128 \times 3$. The preprocessed driver image provide three channel color signal that are red, green, blue. 

\subsection{AI-based Module Development}
The Drive safe platform consists of several AI-based modules. In particular, These modules include capsule network, convolutional neural network, Apriori, and Bayesian network (BN) for driver activity recognition, driver mood mining, sequential cognitive pattern mining, and driver mood repairment content recommender, respectively. A combined effort of those AI modules fulfill the goal of intelligent transportation cyber-physical system towards the cognitive-behavioral mining of the vehicle drive. In particular, a personalized safe driving cyber-physical system is imposed that reduces the risk of road accident due to the distracted driving.  

\begin{figure*}[!t]
%	\centering{\includegraphics[width=\textwidth,height=6.5cm]{fig/Capsule_Activity.pdf}}
\centering{\includegraphics[width=\textwidth]{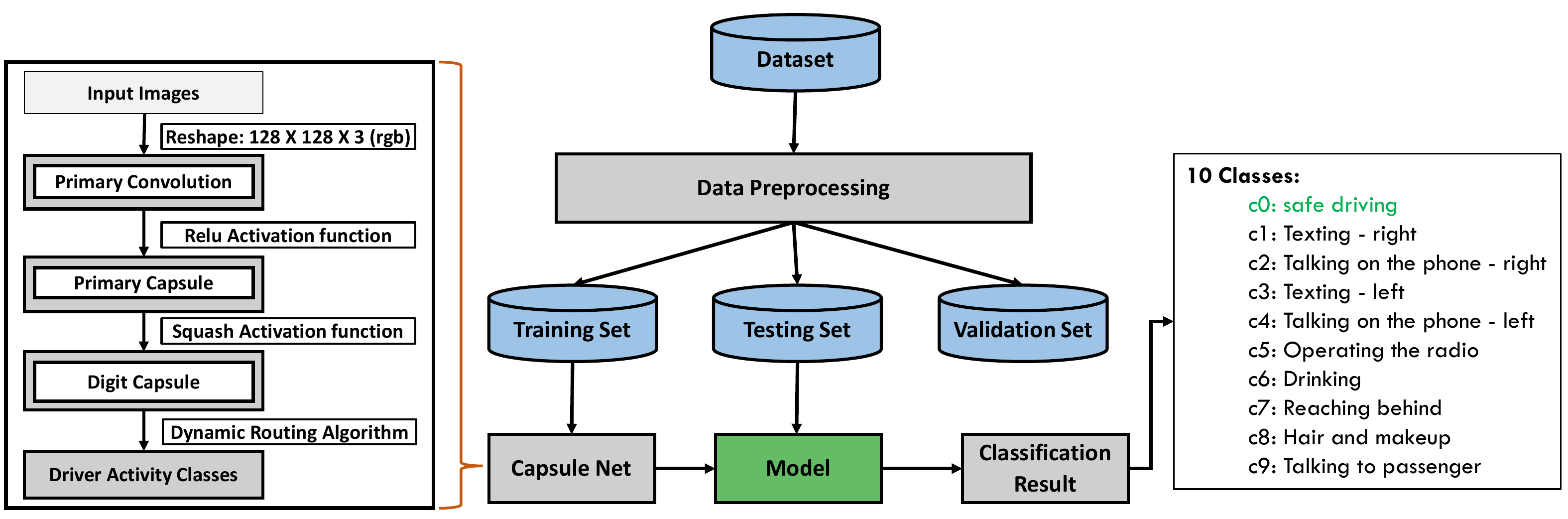}}
	\caption{Capsule network design for driver activity recognition.}
	\label{Capsule_Activity}
\end{figure*}

\subsubsection{Driver Activity Recognition}
In order to driver activity recognition, we consider per-trained driver activity AI model. That model trained by capsule network \cite{sabour2017dynamic} using state of-the-art state farm distracted driver detection dataset \cite{driverdataset}. This dataset consists of ten different drive activities during the driving as shown in Table \ref{tab:tab_driver_activity_class} and two major classes: safe driving, and distracted driving. Additionally, we have used all ten classes (seen in Table \ref{tab:tab_driver_activity_class}) to fuse the driver activity with mental state for the drive safe platform in IT-CPS.

\begin{table}[!h]
	\caption{Capsule network parameters \cite{munir2019towards}}
	\begin{center}
		\begin{tabular}{|c|c|c|}
			\hline
			\textbf{Parameter}&\textbf{Primary convolution}&{\textbf{Primary capsule}} \\
			\hline
			Filters/Capsules & $256$ & $32$ \\
			\hline
			kernel & $7$ & $9$\\
			\hline
			Strides & $1$ & $2$\\
			\hline
			Dimension of capsule & x & $8$\\
			\hline
		\end{tabular}
		\label{tab:capsuple_config}
	\end{center}
\end{table}

To train the pre-trained driver activity recognition model, we stratify the dataset \cite{driverdataset} as a distribution of each class in train, test and validation approximately in equal quantity. Thus, we consider training, validation, and testing samples $14400$, $4000$, and $4000$, respectively, during the pre-trained CN model training process. The drivers physical activity during the driving relies on sharp changes of the body parts; especially, on the angle of the movement of face, rest, nick, etc. Thus, CN \cite{sabour2017dynamic} can handle those features of the body parts that is caused by sharp movement better than the CNN \cite{xing2019driver} due to its pooling mechanism. The driver activity recognition pre-trained model is depicted in Fig. \ref{Capsule_Activity} and a major configuration parameters of the CN model is shown in Table \ref{tab:capsuple_config}.

\begin{figure}[!h]
	\centering
	\includegraphics[width=8.0cm,height=6.0cm]{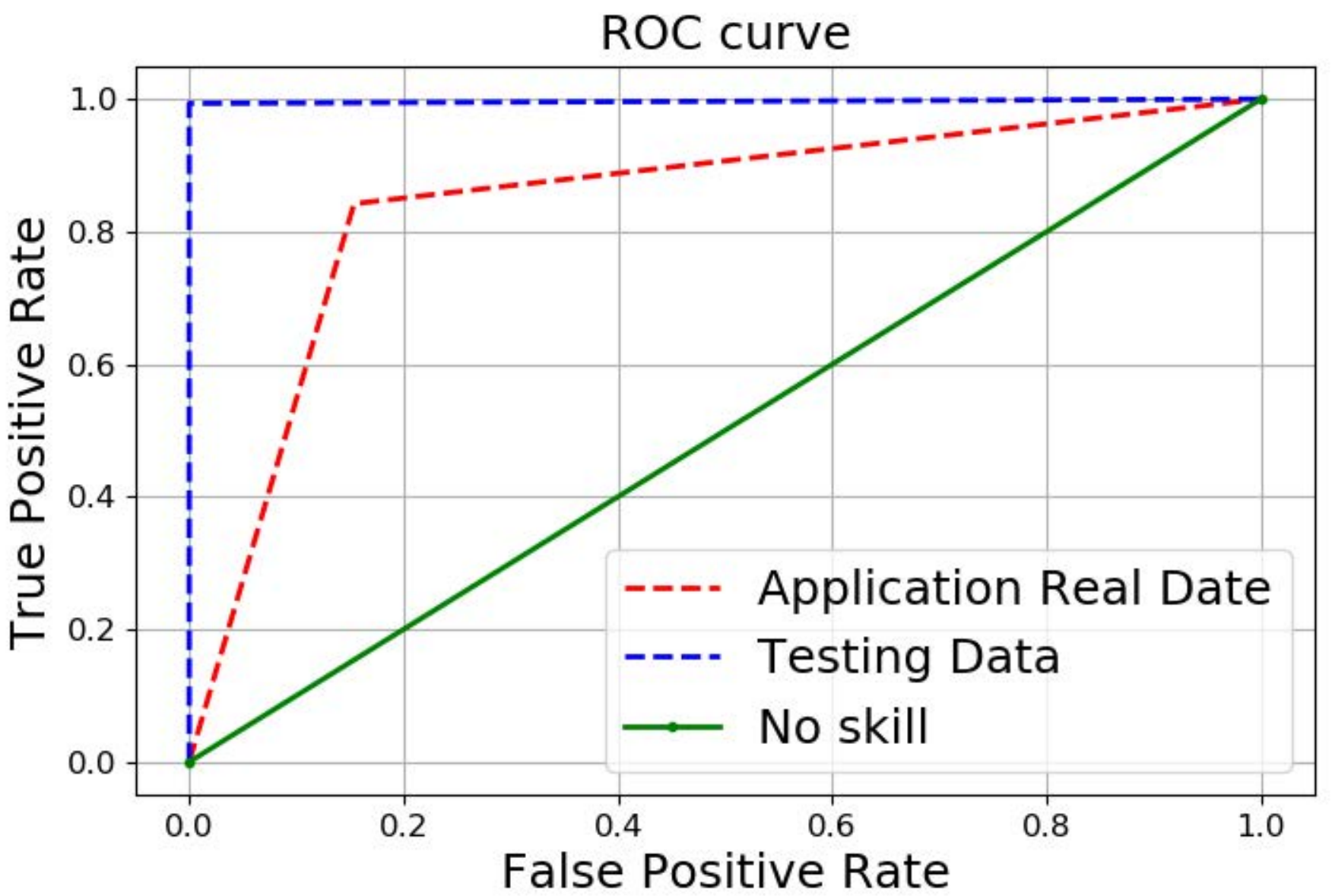}
	\caption{Receiver operating characteristic (ROC) curve for driver activity recognition.}
	\label{Roc_cn}
\end{figure}

\begin{figure}[!h]
	\centering
	\includegraphics[width=8.0cm,height=6.0cm]{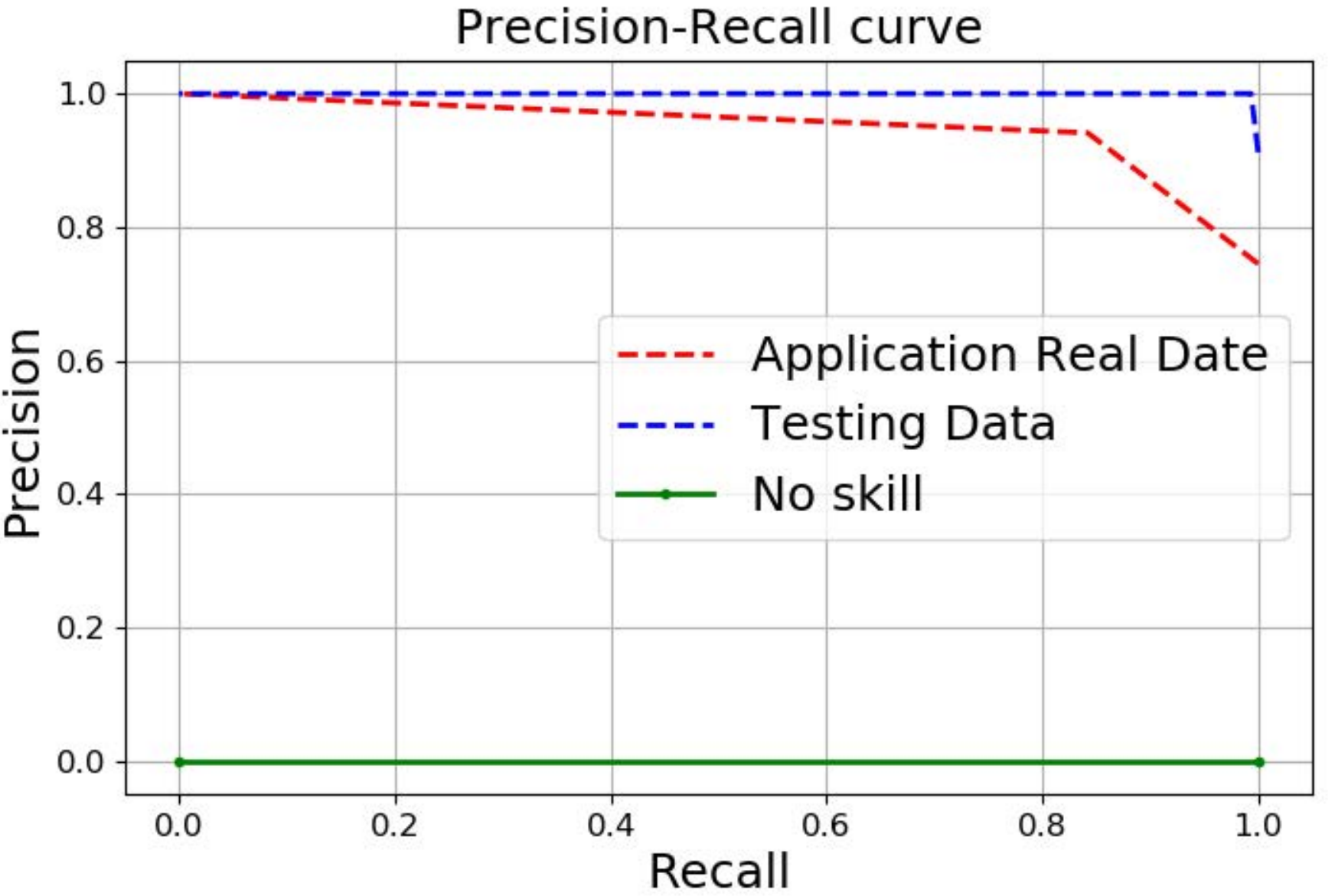}
	\caption{Precision-recall curve for driver activity recognition.}
	\label{recall_cn}
\end{figure}

\begin{table}[!h]
	\caption{Activity classification report for real-time driver data}
	\begin{center}
		\begin{tabular}{|c|c|c|c|c|}
			\hline
			\textbf{Name}&\textbf{Precision}&\textbf{Recall}&\textbf{F1-score} &\textbf{Support}\\
			\hline
			Safe driving & $0.65$ & $0.85$ & $0.73$ & $1300$\\
			\hline
			Distracted driving & $0.94$ & $0.84$ & $0.89$ & $3800$\\
			\hline
%			Micro avg & $0.84$ & $0.84$ & $0.84$ & $5100$\\
%			\hline
			Micro avg & $0.79$ & $0.84$ & $0.81$ & $5100$\\
			\hline
			Weighted avg & $0.87$ & $0.84$ & $0.85$ & $5100$\\
			\hline
		\end{tabular}
		\label{tab:classification_report_real_time}
	\end{center}
\end{table}

\begin{table}[!h]
	\caption{Activity classification report using test set}
	\begin{center}
		\begin{tabular}{|c|c|c|c|c|}
			\hline
			\textbf{Name}&\textbf{Precision}&\textbf{Recall}&\textbf{F1-score} &\textbf{Support}\\
			\hline
			Safe driving & $0.94$ & $1.00$ & $0.97$ & $88$\\
			\hline
			Distracted driving & $1.00$ & $0.99$ & $1.00$ & $912$\\
%			\hline
%			Micro avg & $0.99$ & $0.99$ & $0.99$ & $1000$\\
			\hline
			Micro avg & $0.97$ & $1.00$ & $0.98$ & $1000$\\
			\hline
			Weighted avg & $0.99$ & $0.99$ & $0.99$ & $1000$\\
			\hline
		\end{tabular}
		\label{tab:classification_report_training}
	\end{center}
\end{table}

We have exported and deploy the pre-trained driver activity recognition model on-vehicle dashboard client application. Therefore, we provide driver activity recognition feedback in a $300$ ms interval to driver via the developed on-vehicle dashboard client application. Meanwhile, we sent the activity ID for the further fusion with the driver mood. Receiver operating characteristic (ROC) and Precision-recall curves are illustrated in Figs. \ref{Roc_cn}, and \ref{recall_cn}, respectively, for the driver activity recognition. In addition, the real-time and training of driver activity classification report are given in Tables \ref{tab:classification_report_real_time} and \ref{tab:classification_report_training}, respectively. The accuracy of the real-time driver activity recognition is around $84\%$ due to the different camera position and vehicle jerking.       

\subsubsection{Driver Mood Mining}
The goal of a vehicle driver mood mining is to find a cognitive relationship based on the physiological sensory observations. In particular, by extracting features of physiological observations that are captured by EMG, ECG, EDA, and EEG sensors from the vehicle drive BAN. These observations will be classified into $1$ to $9$ for the arousal and valence values \cite{posner2005circumplex,bradley2000affective}. A relationship among the driver mood, arousal, and valence scores are shown in Table \ref{tab:mood_arousal_valence}.

\begin{table*}[!h]
	\caption{Driver mood category \cite{scherer2005emotions}}
	\begin{center}
		\begin{tabular}{|c|c|c||c|c|c||c|c|c|}
			\hline
			\textbf{Valence}&{\textbf{Arousal}} & \textbf{Mood}& \textbf{Valence}&{\textbf{Arousal}} & \textbf{Mood}& \textbf{Valence}&{\textbf{Arousal}} & \textbf{Mood}\\
			\hline
			$1$ & $1$ &Dejected & $2$ & $1$ & Dejected & $3$ & $1$ &Doubtful \\
			\hline
			$1$ & $2$ &Wavering & $2$ & $2$ & Wavering & $3$ & $2$ &Anxious \\
			\hline
			$1$ & $3$ &Gloomy & $2$ & $3$ & Depressed & $3$ & $3$ &Ashamed \\
			\hline
			$1$ & $4$ &Gloomy & $2$ & $4$ & Uncomfortable & $3$ & $4$ &Taken back \\
			\hline
			$1$ & $5$ &Disappointed & $2$ & $5$ & Startled & $3$ & $5$ &Distrustful\\
			\hline
			$1$ & $6$ &Bitter& $2$ & $6$ & Insulted & $3$ & $6$ &Frustrated \\
			\hline
			$1$ & $7$ &Loathing & $2$ & $7$ & Disgusted & $3$ & $7$ &Distressed \\
			\hline
			$1$ & $8$ &Contemptuous & $2$ & $8$ & Defiant & $3$ & $8$ &Hateful \\
			\hline
			$1$ & $9$ &Obstructive & $2$ & $9$ & Obstructive & $3$ & $9$ &Angry \\
			\hline
			$4$ & $1$ & Droopy & $5$ & $1$ &Tired & $6$ & $1$ & Sleepy \\
			\hline
			$4$ & $2$ & Bored & $5$ & $2$ &Melancholic & $6$ & $2$ & Conscientious\\
			\hline
			$4$ & $3$ & Embarrassed & $5$ & $3$ &Languid & $6$ & $3$ & Longing \\
			\hline
			$4$ & $4$ & Feel guilt & $5$ & $4$ &Worried & $6$ & $4$ & Confident \\
			\hline
			$4$ & $5$ & Apathetic & $5$ & $5$ &Neutral & $6$ & $5$ & Impressed \\
			\hline
			$4$ & $6$ & Suspicious & $5$ & $6$ &Impatient & $6$ & $6$ & Passionate \\
			\hline
			$4$ & $7$ & Indignant & $5$ & $7$ &Jealous & $6$ & $7$ & Feeling superior \\
			\hline
			$4$ & $8$ & Annoyed & $5$ & $8$ &Afraid & $6$ & $8$ & Conceited \\
			\hline
			$4$ & $9$ & Hostile & $5$ & $9$ &Tensed & $6$ & $9$ & Lusting \\
			\hline
			$7$ & $1$ &Servient & $8$ & $1$ & Compassionate & $9$ & $1$ &Conductive \\
			\hline
			$7$ & $2$ &Polite & $8$ & $2$ & Peaceful & $9$ & $2$ & Calm \\
			\hline
			$7$ & $3$ &Contemplative & $8$ & $3$ & Friendly & $9$ & $3$ &Serene  \\
			\hline
			$7$ & $4$ &Attentive & $8$ & $4$ & Hopeful & $9$ & $4$ & Glad \\
			\hline
			$7$ & $5$ &Interested & $8$ & $5$ & Feel Well & $9$ & $5$ & Pleased \\
			\hline
			$7$ & $6$ &Amused & $8$ & $6$ & Determined & $9$ & $6$ & Happy \\
			\hline
			$7$ & $7$ & Convinced & $8$ & $7$ & Enthusiastic & $9$ & $7$ &Enthusiastic \\
			\hline
			$7$ & $8$ &Ambitious & $8$ & $8$ & Excited & $9$ & $8$ & Courageous \\
			\hline
			$7$ & $9$ & Aroused/Astonied & $8$ & $9$ & Adventurous & $9$ & $9$ & Self confident \\
			\hline
		\end{tabular}
		\label{tab:mood_arousal_valence}
	\end{center}
\end{table*}

\begin{figure}[!h]
	\centering
	\includegraphics[width=8.7cm,height=5.5cm]{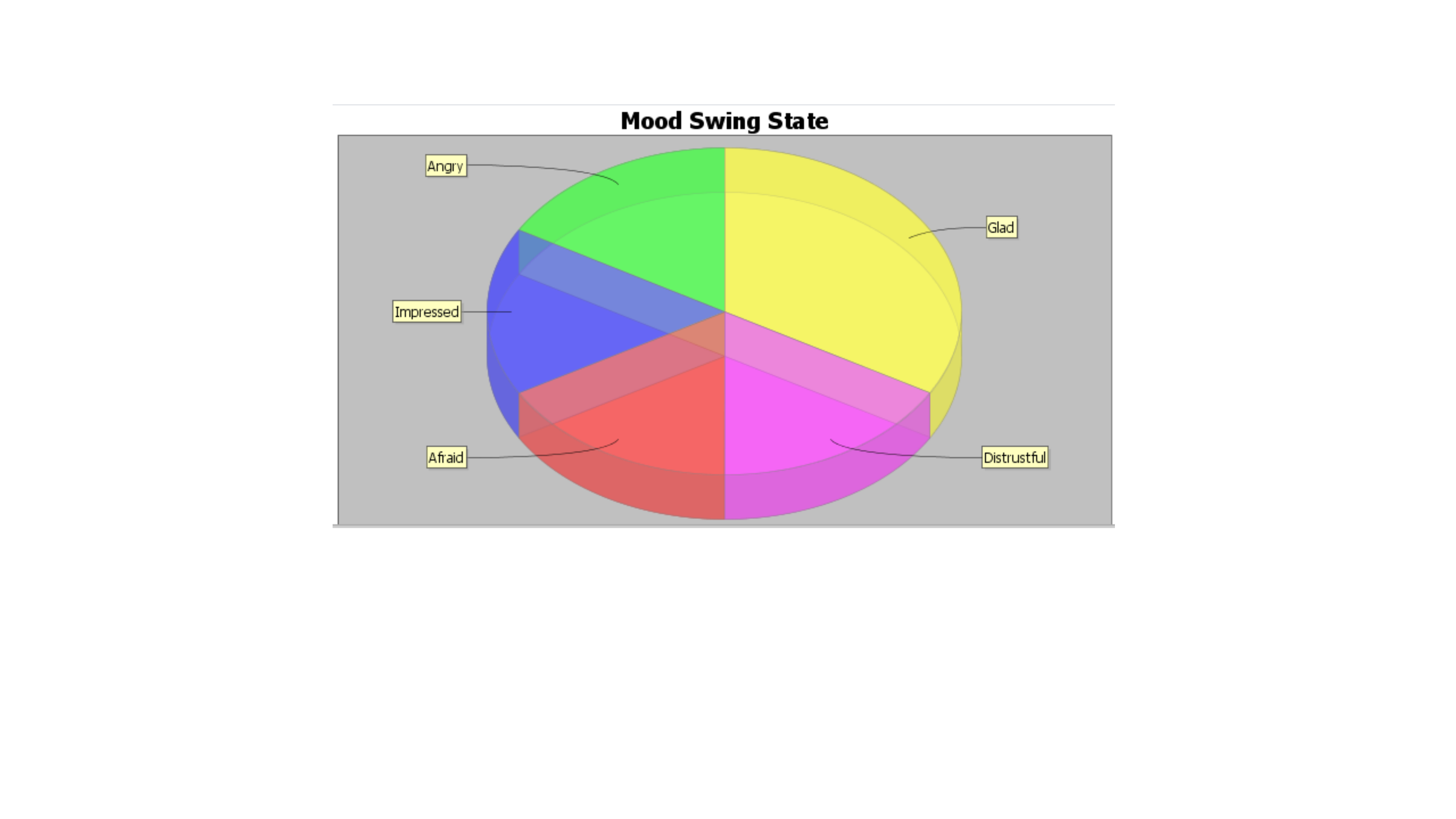}
	\caption{CNN-based emotion wheel of personalized vehicle driver.}
	\label{Simulation_mode_swing}
\end{figure} 

In this work, we deploy a learning model that can effectively find the arousal and valence from the driver's biosensor data. Therefore, we have used our CNN-based  
\cite{alam2019healthcare} mood mining framework that can effectively determine arousal and valence for the driver. In addition, the considered CNN-based mood mining model is trained by the DEAP dataset \cite{koelstra2011deap}. In this work, we feed an additional EEG sensor that can capture the brain activity of the vehicle driver during the driving. Thus, in an observational period, we feed $8064$ data points for each EMG, ECG, EDA, and EEG sensors and examine in the edge server. The output of this CNN model consists of a pair of integer values that are valence and arousal. In order to cognitive-behavioral mining of the driver during the driving, we fuse driver physical activity with the mental state by applying the sequential pattern mining mechanism. That can infer a relationship between the mental and physical activity of the vehicle driver toward a safe driving event. Fig. \ref{Simulation_mode_swing} depicts the output of the mood mining of a particular driver.       

\subsubsection{Sequential Pattern Mining}
\begin{figure}[!h]
	\centering
	\includegraphics[width=8.0cm,height=6.0cm]{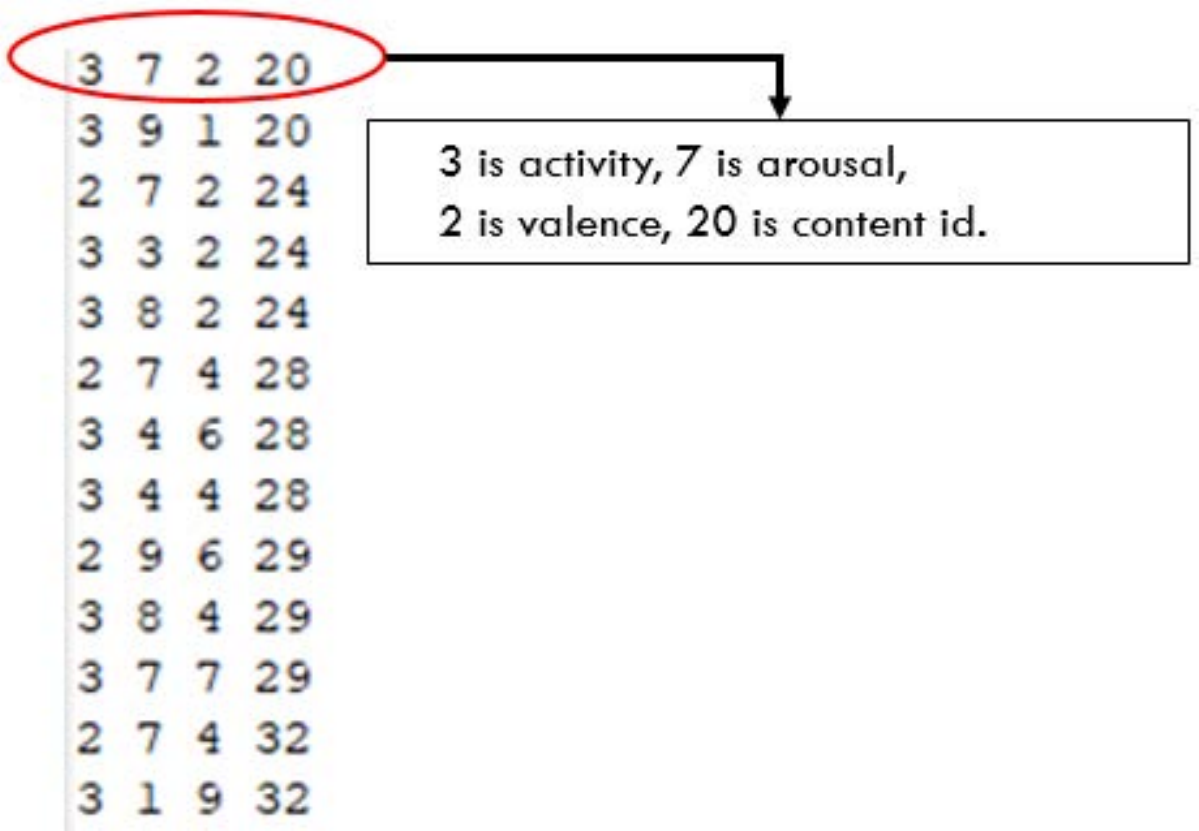}
	\caption{Apriori transaction example.}
	\label{Apriori_Transaction}
\end{figure}
The driver activity, on-compartment vehicle environmental likelihood and driver mood (i.e., valence and arousal) are fused with the audio content in order to generate the cognitive-behavioral pattern of the vehicle driver status. In particular, this cognitive-behavioral sequential pattern map with the audio content candidate for the driver mood repairment during the drive. Thus, we have developed association rule mining by following the concept of the Apriori \cite{agrawal1993mining} model. Through the association rule mining, the user behavioral pattern is generating so that the system can recommend a candidate set of contents based on the contextual information. 

In the association rule mining, each of the patterns is considered as transactions (as seen in Fig. \ref{Apriori_Transaction}), and based on the confidence and support value the fused data is sent to the emotion aware recommender. In this pattern mining method, we consider minimum support count $0.1$ so that we can capture the significant patterns among the physical activity and driver's physiological behavior. In particular, we determine a sequence of tuple $\langle {activity}_3, {arousal}_7, {valence}_2, {content}_{20} \rangle$ that provides a set of rules based on minimum support count and minimum confidence value. In other words, we generate association rules that represents a mapping $\langle {activity}_3, {arousal}_7, {valence}_2 \rangle \to {content}_{20}$ between the cognitive behavior of the driver and candidate content. We employ this sequential pattern rules to a Bayesian network that can recommend an appropriate content towards the driver mood repairment.

\subsubsection{Content Recommendation for Mood Repairment}
\begin{figure}[!h]
	\centering
	\includegraphics[scale=.5]{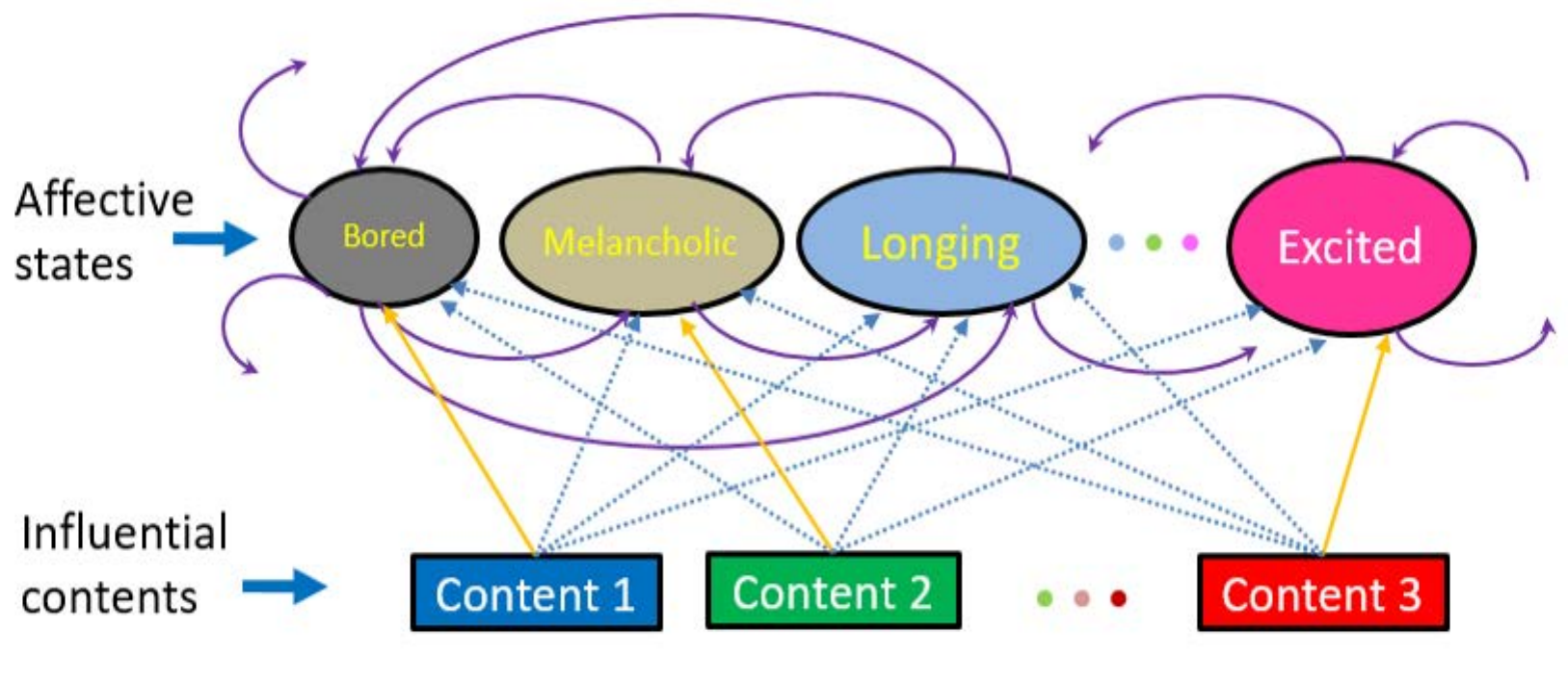}
	\caption{Bayesian interface model for content recommendation.}
	\label{Bayesian}
\end{figure}

One of the key objectives of our proposed drive safe platform is the recommendation of content for the mood repairment of the vehicle driver. To meet the goal, we design a Bayesian inference based recommendation system that can recommend the content to the vehicle driver. A Bayesian network \cite{friedman1997bayesian} is a probabilistic graphical model that represents a set of random variables and their conditional dependencies via a graph. The hidden affective states and the contents can be represented as the Bayesian network as shown in Fig. \ref{Bayesian}.

A set $\left\{{Bored, Melancholic, Longing, \dots, Excited}\right\}$ of $N$ states can be mathematically represented as $\mathcal{S} =\left\{{s_1, s_2, s_3, \dots, s_N}\right\}$. The contents are $\mathcal{S} =\left\{{Content}_1, {Content}_2, \dots, {Content}_N\right\}$, which can be mathematically represented as $\mathcal{C} =\left\{{c_1, c_2, \dots, c_N}\right\}$. The Bayesian inference \cite{friedman1997bayesian} is a method of statistical inference in which Bayes' theorem is used. That can update the probability for a hypothesis as more evidence or information becomes available. Using the Bayesian theory, we can determine the probability of future affective states based on the stimuli content. We consider current affective state $P(s_i | c_j, s_{i-1})$, where $s_i$ is the future affective state, $c_j$ represents current audio content (listing), and  $s_{i-1}$ denotes the current affective state. However, our goal is to find out a sequence of the content list for smooth mood swing from negative to positive affects. Such sequence, we have determined through maximum likelihood estimation through Bayesian estimator like the Viterbi algorithm \cite{forney1973viterbi}. In order to operate the proposed intelligent drive safe platform, we have developed an on-vehicle dashboard as a client application. That can effectively and autonomously provide feedback to the vehicle drive while driving.    

\subsection{On-vehicle Dashboard}
\begin{figure*}[!h]
	\captionsetup[subfigure]{justification=centering}
	\centering
	\begin{subfigure}{0.45\textwidth}
		\includegraphics[width=\textwidth]{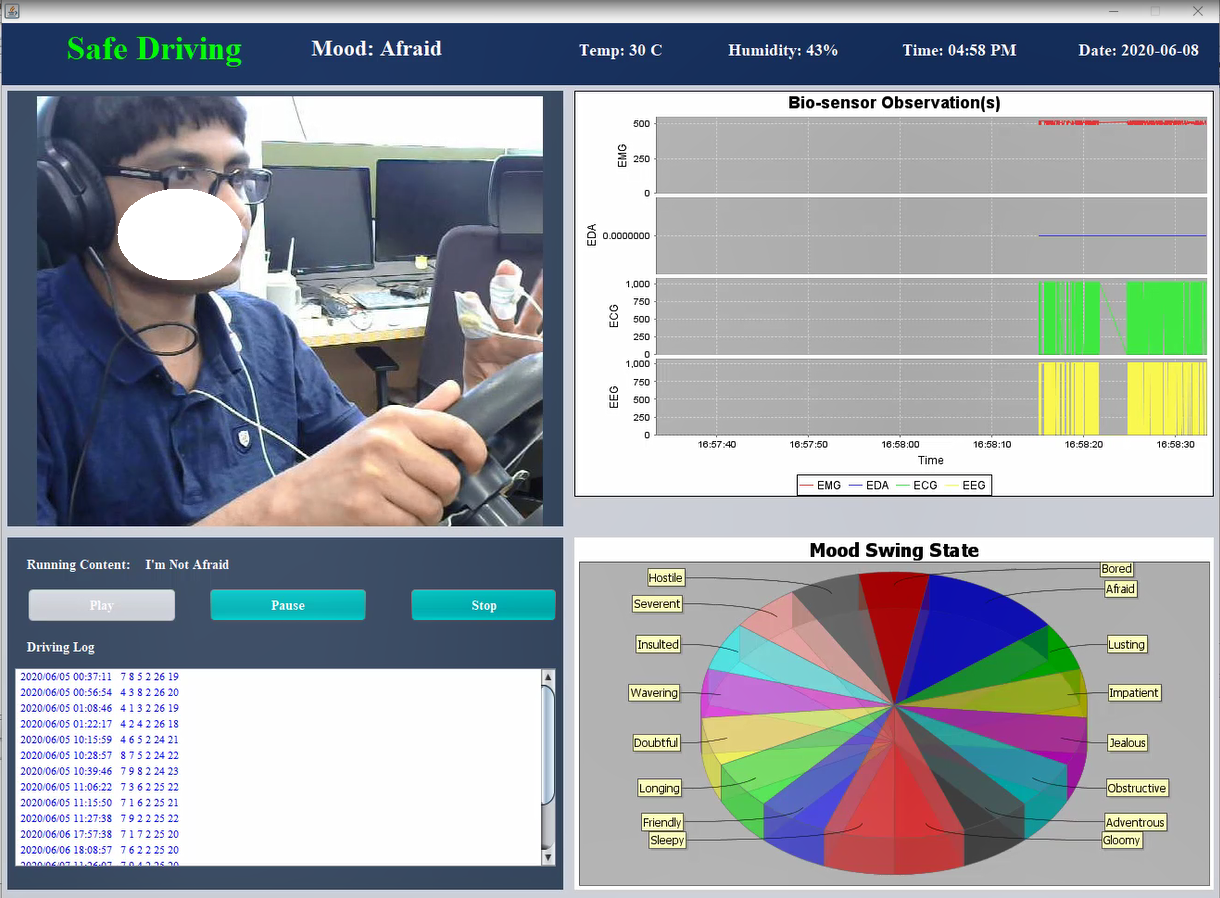}
		\caption{Safe driving with all features.}
		\label{fig:Image_1}
	\end{subfigure}
	\begin{subfigure}{0.45\textwidth}
		\includegraphics[width=\textwidth]{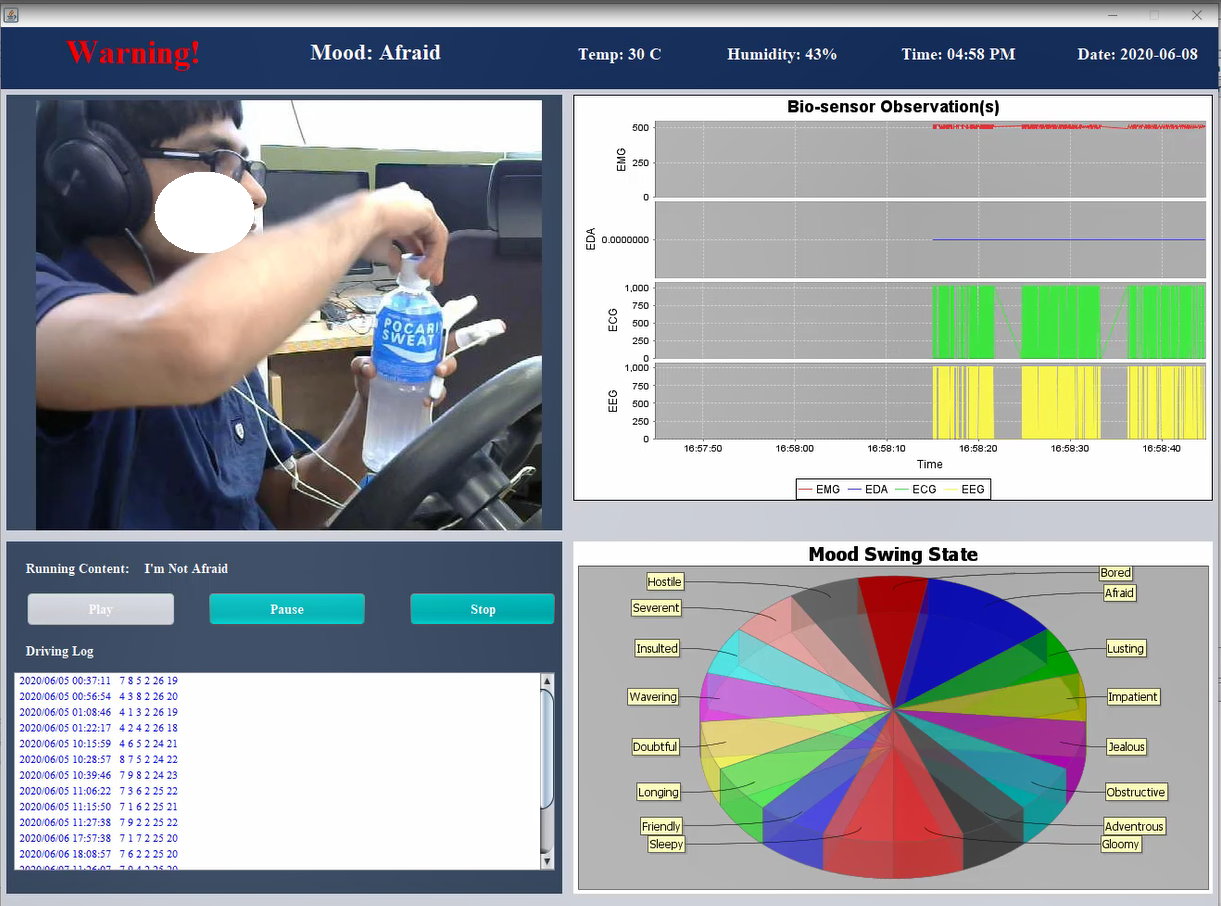}
		\caption{Distracted driving with all features.}
		\label{fig:Image_2}
	\end{subfigure}
	\begin{subfigure}{0.45\textwidth}
		\includegraphics[width=\textwidth]{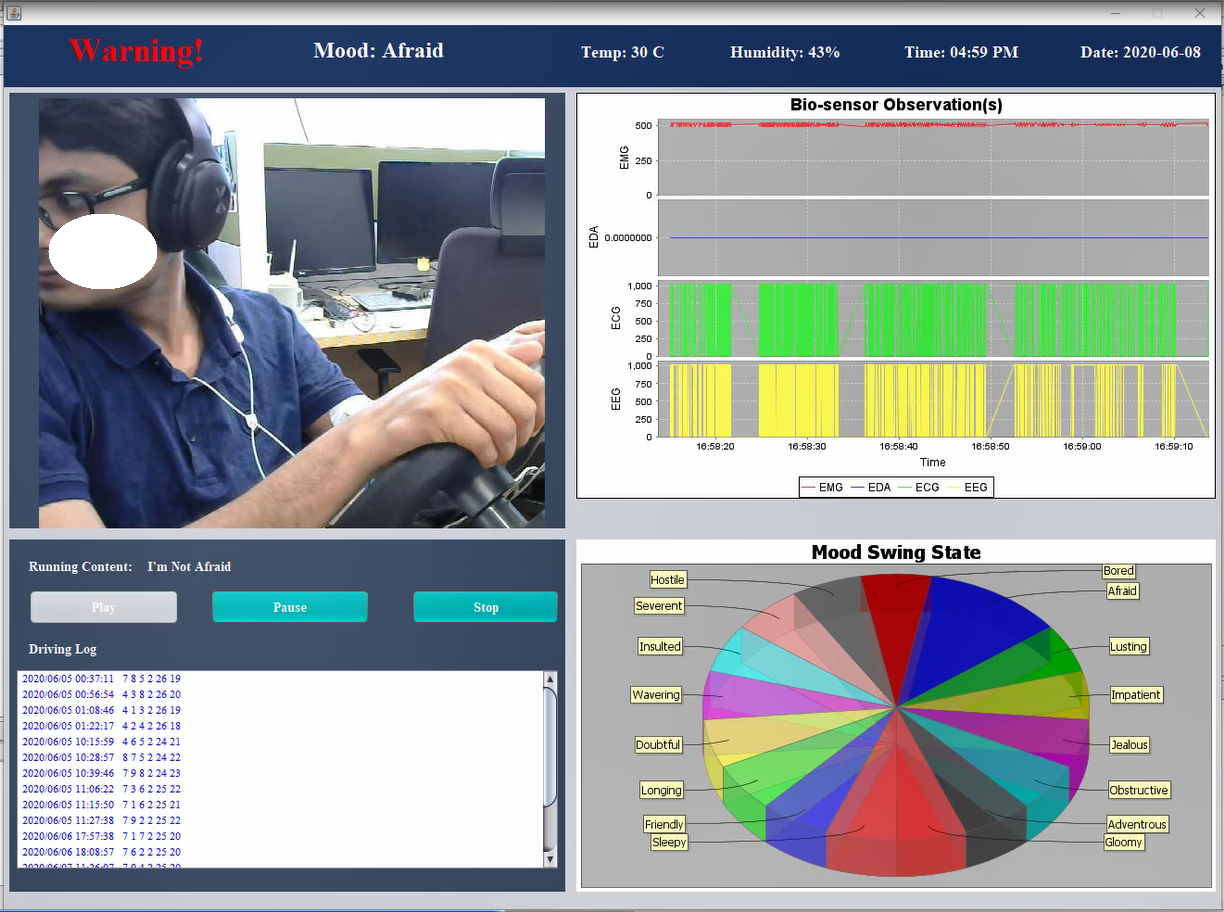}
		\caption{Distracted driving with the previous mood.}
		\label{fig:Image_5}
	\end{subfigure}
	\begin{subfigure}{0.45\textwidth}
		\includegraphics[width=\textwidth]{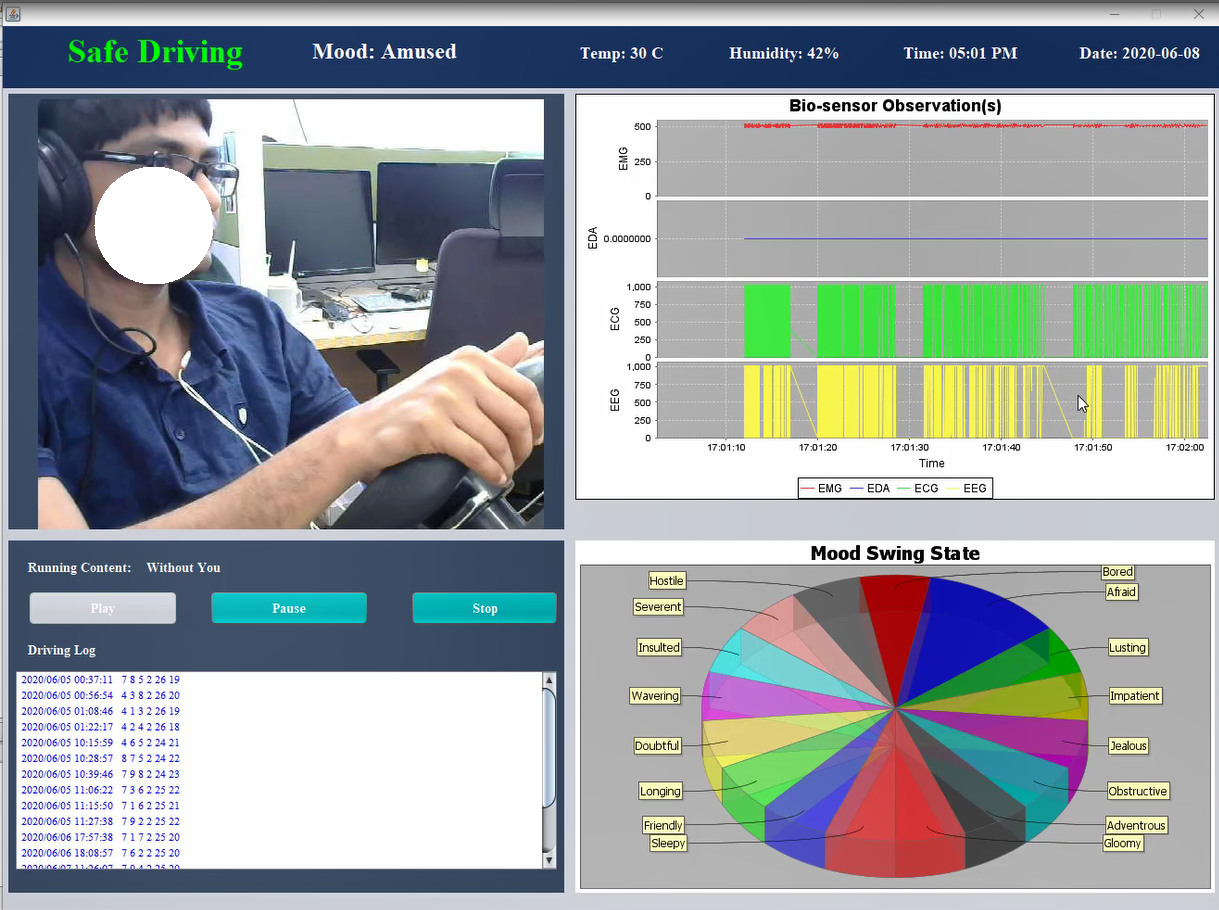}
		\caption{Audio changes with mood changed.}
		\label{fig:Image_6}
	\end{subfigure}
	\label{Vehicle_dashboard}
	\caption{Evaluation environment of vehicle dashboard.}
\end{figure*}

The role of the developed on-vehicle dashboard is to guide the vehicle driver from the distracted driving. In particular, this dashboard application can operate without the interaction of the vehicle driver during the driving while autonomously playing the recommended audio content for the driver's affective state. Further, notify the driver based on on-compartment activity toward the drive safe. We demonstrate the developed on-vehicle dashboard client application of the drive safe platform in Figs. \ref{fig:Image_1}, \ref{fig:Image_2}, \ref{fig:Image_5}, and \ref{fig:Image_6}. This dashboard also visualizes the mood swing state that assures the driver's lifelog.

To developed the on-vehicle dashboard UX, we have followed the \emph{recognition rather than recall} NIELSEN’S heuristic \cite{10_Usability} UX design principle. In which, the drivers can easily detect their current mood when they watch the dashboard. They actually use the recognition usability heuristic as all the cues are visible in the dashboard. For example, when the drivers’ mood is changed, the UI automatically detects his current status. After that, it can change the audio based on the drivers’ mood. In the UI, the driver can also come to know that either the driving is safe or there is a warning during driving by watching the dashboard. The vehicle driver can easily understand the situation with the clear interaction of the developed on-vehicle dashboard and does not need to rethink it. Apart from that, the developed on-vehicle dashboard for driver's affective mood repairment UX satisfies the other NIELSEN'S usability heuristics \cite{10_Usability} and a relationship between those usability heuristics with the developed prototype are described in Table \ref{tab:usability}. Additionally, the output of this on-vehicle dashboard ensures a personalized recommended system so that this dive safe platform maintain the driver's personal data privacy. 

\begin{table*}[!t]
	\caption{NIELSEN'S usability heuristics meet with the developed driver dashboard UX }
	\centering
	\begin{tabular}{|p{3.2cm}|c|p{11.5cm}|}
		\hline
		\textbf{Heuristic Name} & \textbf{Meets/Not?} & \textbf{Comments}\\ 
		\hline
		Visibility of system status &  Yes  & The on-vehicle dashboard can interact with users by always visible and visual feedback.User can recognize the system is working or not by using precise and reasonable feedback. \\ 
		\hline
		Match between system and the real world & Yes & The on-vehicle dashboard prototype can interact with vehicle driver and internal system functions, such as activity recognition, content recommender, and so on. \\ 
		\hline
		Flexibility and efficiency of use & Yes & The developed prototype is flexible that helps drivers to use in an efficient way as expert users' and also friendly to the new system user. \\ 
		\hline
		Consistency and standards & Yes & Usages of the on-vehicle dashboard are predictable and learnable to the vehicle driver due to interactive follow of functionality.\\ 
		\hline
		Error Prevention & Yes & Provides proper feedback messages to users.\\ 
		\hline
		Aesthetic and Minimalist Design & Yes & Lifelog and visualization of the driver mood fluctuation are keeping the content for communicative to the driver. \\ 
		\hline
		Help users recognize, diagnose, and recover from errors & No & Will be added in the future works. \\ 
		\hline
		Help and documentation & No & Will be added in the future works. \\ 
		\hline
		User control and freedom & No & Will be added in the future works. \\ 
		\hline
	\end{tabular}
	\label{tab:usability}
\end{table*}

\begin{figure}[!h]
	\centering
	\includegraphics[width=8.6cm,height=9.5cm]{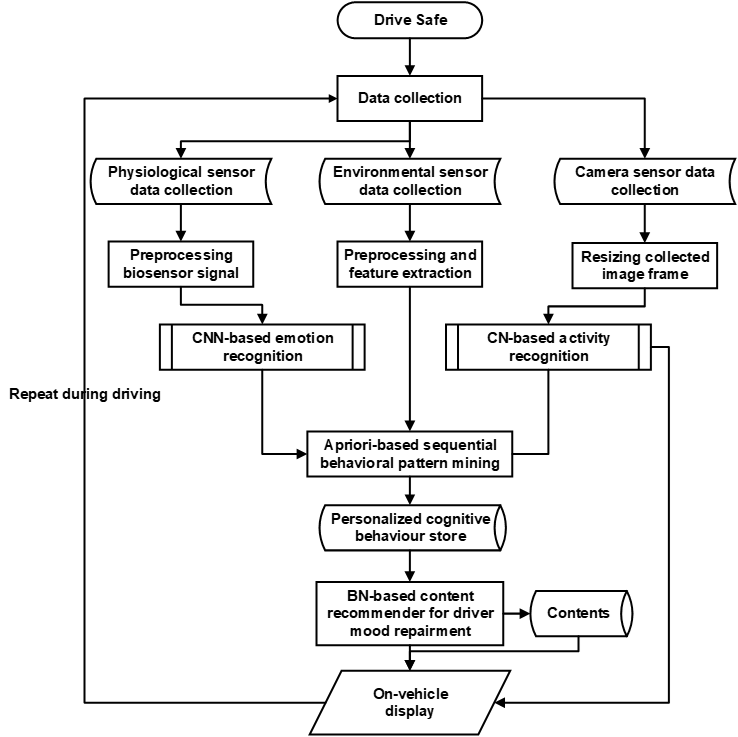}
	\caption{Drive safe platform algorithmic procedure in intelligent transportation cyber-physical system.}
	\label{overall_process}
\end{figure}
The drive safe platform algorithmic procedure illustrates in Fig. \ref{overall_process}. The entire procedure not only meets the requirement of pervasive computing with the IT-CPS but also ensure the road safety during the drive. A brief discussion and key findings of the drive safe platform are described in the later section.

\section{Experimental Discussion and Key Findings}
\label{sec:discussion}
\begin{table*}[!h]
	\caption{Summary of the technologies and methodologies for the proposed cognitive-behavioral mining in intelligent transportation cyber-physical system}
	\begin{center}
		\begin{tabular}{|c|c|c|}
			\hline
			\textbf{Purpose}&\textbf{Method}&\textbf{ Technology}\\
			\hline
			Driver activity recognition & Capsule network & Deep neural network-based AI model \\
			\hline
			Driver mood mining & Convolutional neural network & Deep neural network-based AI model \\
			\hline
			Environmental features analysis for fusion & Maximum likelihood probability model & Statistical model \\
			\hline
			Sequential behavioural pattern mining & Apriori algorithm & Intelligent pattern mining model \\
			\hline
			Content recommendation for affective mood repairment & Bayesian network & Statistical AI model \\
			\hline
			On-vehicle dashboard application & Recognition rather than the recall & Heuristics for UI design \\
			\hline
			Model computation & N/A & Multi-access edge computing and on-device \\
			\hline
			Communication & Multi-thread asynchronous & Bluetooth, WiFi, Wired (LAN) \\
			\hline
		\end{tabular}
		\label{tab:technology_mapping}
	\end{center}
\end{table*}
The goal of this work is to provide a proof of concept (PoC) for the cognitive-behavioral mining of the vehicle driver in intelligent transportation cyber-physical system. Therefore, we have employed five AI and statistical-based models to infer the cognitive-behavioral mining of a vehicle driver that can ensure safe driving during the drive. Thus, in this paper, the performance analysis of each individual model is out of the scope. 
Hence, we have performed a qualitative evaluation for the developed drive safe platform. Thus, in this section, first, we will discuss statistical analysis, ANOVA test, and confidence interval analysis based on user study for the developed drive safe platform. Then we will provide some of the key findings and technical challenges in terms of prototype development for the drive safe platform in intelligent transportation cyber-physical system.

\subsection{Qualitative Evaluation of Drive Safe}
\subsubsection{Experimentation Design}
In this section, we discuss a detailed procedure for evaluation the developed drive safe prototype of the IT-CPS.
\paragraph{Population}
In this experiment, we consider five participants for testing the usability and effectiveness of the developed drive safe prototype of the IT-CPS. We have not divided the participants into groups since all of the users of the developed system are considered as novices. The developed drive safe prototype of the IT-CPS is considered as a new filed. Therefore, most of the users are not experts.    
\paragraph{Hypotheses}
In this work, the primary hypotheses are to operate the developed drive safe prototype of the IT-CPS without any physical interaction (i.e., without input from the user) by the vehicle driver during the driving. Meanwhile, the system should autonomously play the recommended audio content for the driver’s affective state and notifying the driver based on on-compartment activity to ensure safe driving. Secondary hypotheses are considered to the visualization of the current activity, mood swing state, and other sensory data (i.e., physiological sensors, camera sensors, and environmental sensors), as well as the driver’s lifelog. 
\paragraph{Study Conditions}
In this experiment, our goal is to conduct a one-way within-subject design to evaluate the developed UX prototype. In which, we design six individual questions based on NIELSEN’S usability heuristics \cite{10_Usability} to perform the usability test of the developed drive safe prototype of the IT-CPS. Therefore, we consider independent variable based on the score value of each question and all six questions will be scored by the participants ([See Appendix \ref{apx:Observational_questionnaires}]). The score level are as follows, 1: Not good, 2: Somehow good, 3: Good, 4: Satisfactory, 5: Very satisfactory. Furthermore, in order to evaluate the statistical confidence level based on the user feedback, we design another eight questions ([See Appendix \ref{apx:overall_questionnaires}] with a binary level, 0: No, 1: Yes.

\subsubsection{Experimental Results}
\begin{figure}[!h]
	\centering
	\includegraphics[scale=0.5]{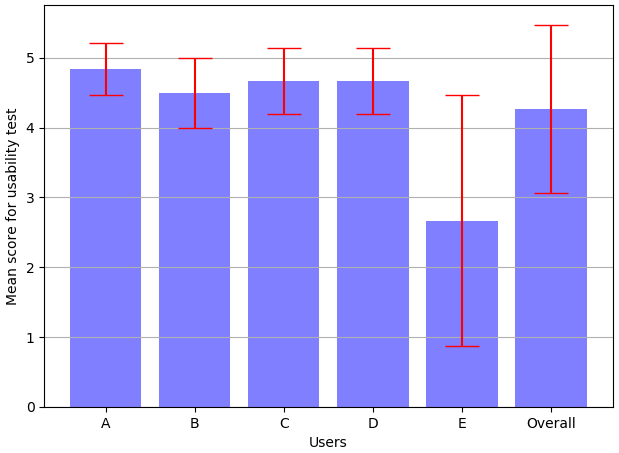}
	\caption{Standard deviation analysis for the users feedback using error bar analysis.}
	\label{error_bar}
\end{figure}

We analyze the users evaluation data on python platform. We have performed several statistical analysis using the users feedback data to evaluate the developed drive safe prototype of IT-CPS. In particular, we have performed statistical analysis (i.e., mean, variance, and standard deviation), ANOVA test, and confidence interval analysis using five well-known methods. An error bar from the standard deviation and mean is represented in Fig. \ref{error_bar}. Meanwhile, we describe the mean, variance, and standard deviation of the usability test scores in Table \ref{tab:mean}. Further, we have conducted a one-way analysis of variance (ANOVA) based on the user feedback scores of the usability test of the save drive prototype. Table \ref{tab:ANOVA} illustrates the outcome of that test. In this ANOVA test, we have found the P-value as $0.0041$ that is less than the $0.05$. Therefore, our usability test achieve the statistical significance for the conducted experiment. In case of the F-value (i.e., $5.02$), we can say the variance between the means of among populations significantly different. In other words, the considered hypotheses have the statistical significance that can evaluate the developed drive safe prototype of IT-CPS with a high reliability. 

\begin{table}[!h]
	\caption{Mean, Variance, and Standard deviation of the usability test scores}
	\begin{center}
		\begin{tabular}{|c|c|c|c|}
			\hline
			\textbf{Users/Overall}&\textbf{Mean}&\textbf{Variance}&\textbf{Standard Deviation} \\
			\hline
			A	& $4.83$ & $0.17$ & $0.37$ \\
			\hline
			B	& $4.5$ & $0.3$ & $0.5$ \\
			\hline
			C	& $4.67$ & $0.27$ & $0.47$ \\
			\hline
			D	& $4.67$ & $0.27$ & $0.47$ \\
			\hline
			E	& $2.66$ & $3.87$ & $1.80$ \\
			\hline
			Overall	& $4.26$ & $1.51$ & $1.20$ \\
			\hline
		\end{tabular}
		\label{tab:mean}
	\end{center}
\end{table}
\begin{table}[!h]
	\caption{ANOVA test result}
	\begin{center}
		\begin{tabular}{|c|c|}
			\hline
			\textbf{Name}&\textbf{Value} \\
			\hline
			Sum of Squares Residual	& $24.33$  \\
			\hline
			Sum of Squares Model	& $ 19.53$  \\
			\hline
			Mean Square Residual	& $0.97$  \\
			\hline
			Mean Square Explained	& $4.88$  \\
			\hline
			F-value	& $5.02$  \\
			\hline
			P-value	& $0.0041$  \\
			\hline
		\end{tabular}
		\label{tab:ANOVA}
	\end{center}
\end{table}
 Finally, we have conducted confidence interval analysis for the effectiveness analysis. In order to this, we have perform five kinds of analysis, these include Asymptotic (Wald), Binomial (Clopper-Pearson), Wilson Score interval, Agresti-Coull (adjusted Wald) interval, and Jeffreys interval \cite{Wald}. We have found $37$ positive out of $40$ samples from the $5$ users. We present our confidence analysis result in Table \ref{tab:confidence}. We can see for an upper $95\%$ confidence level (CL), the average outcome more than $97\%$ that indicates the significance of the developed drive safe prototype of IT-CPS.      
\begin{table}[!t]
	\caption{Confidence interval analysis}
	\begin{center}
		\begin{tabular}{|p{1.2cm}|c|c|c|}
			\hline
			\textbf{Method}&\textbf{Prevalence} &\textbf{Lower $95\%$ CL}&\textbf{Upper $95\%$ CL}\\
			\hline
			Normal approx.	& $0.9250$  & $0.8434$ & $1.0066$\\
			\hline
			Clopper-Pearson exact	& $0.9250$ & $0.7961$ & $0.9843$ \\
			\hline
			Wilson	& $0.9250$ & $0.8014$ & $0.9742$  \\
			\hline
			Jeffreys	& $0.9250$ & $0.8132$ & $0.9784$ \\
			\hline
			Agresti-Coull	& $0.9250$ & $0.7943$ & $0.9812$ \\
			\hline
		\end{tabular}
		\label{tab:confidence}
	\end{center}
\end{table}

In summary, the considered hypotheses have shown the significance of the test design that can efficiently evaluate the usability and effectiveness of the developed drive safe prototype of IT-CPS. In particular, the P-value $0.0041$ (i.e., $< 0.05$) of the ANOVA test ensures that the hypotheses, and developed system are completely aligned with user studies. Further,  in the confidence interval analysis, we have found the prevalence value around $0.93$ for a $95\%$ confidence that assures the effectiveness of the proposed drive safe prototype of IT-CPS. 

\subsection{Key Findings}
Summary of key technical challenges as follows:
\begin{itemize}
	\item First, the data acquisition rate of the considered sensors is varied between each of the sensors as well as the data transfer rate also differ among the three data acquisition IoT networks. As a result, the synchronization of sensors data collection is one of the major challenges. In fact, we can not control the data acquisition rate of the sensors due to the individual characteristics of each sensor. In order to overcome this challenge, we apply a multi-thread asynchronous mechanism that can collect and store sensors data independently.
	\begin{remark}
		The sensors we have used to collect the physiological observations for the drive that can be a commercial wearable product. In fact, that product will reduce the physical hazard for the driver to wear sensors. 
	\end{remark}
	\item Second, in the proposed drive safe platform, the driver activity recognition is one of the crucial requirements to detect the distractions of the vehicle drivers. Therefore, we have deployed a pre-trained driver activity recognition model in the on-vehicle device. The accuracy of that driver distraction detection is around $84\%$ due to the different camera positions for each of the vehicles. Hence, we can improve the model accuracy by deploying an on-vehicle fine tuning model for each vehicle. In that case, the pre-trained driver activity model will act as a generalized model, and each on-vehicle fine tuning model will be a personalized model for each vehicle driver. However, the main drawback of such mechanism is the computational cost for each on-vehicle device, where the energy consumption of the on-vehicle device is proportional to the computational cost. 
	\begin{remark}
		A tradeoff between accuracy and computational cost of the on-vehicle device toward the vehicle driver activity recognition. 
	\end{remark}
	\item Third, the evaluation of driver mood mining from the physiological observations is quite challenging due to the vast sensory observational data generated every two minutes of an observational time period. The two minutes duration is the ideal time to swing the human emotion \cite{posner2005circumplex, bradley2000affective, koelstra2011deap}. Therefore, we send preprocessed data of each physiological observational period to the edge server via the wireless communications along with environmental and activity logs. The driver mood mining is validated by the state-of-the art DEAP dataset \cite{koelstra2011deap} in the edge server to find the arousal and valence for the driver. In fact, another role of the edge server is to calculate a sequential cognitive behavior mining that map with a list of recommended audio contents for the driver's mood repairment. The edge server then only send a list of sequential behavioral pattern mining data to the particular on-vehicle device for personalized decision making.  
	\begin{remark}
		The proposed cognitive-behavioral mining has established multi-access edge computing for intelligent transportation cyber-physical system and ensure the personalized decision for each vehicle driver. 
	\end{remark}
	\item Last but not least, the developed on-vehicle device application can determine a personalized affective state for the vehicle driver by executing a Bayesian network from the lifelog of the individual driver. Further, the on-board vehicle display provides three key functionality, i) safety message to the driver each $300$ ms interval, ii) mood swinging statistics along with lifelog, and iii) the driver's mood repairement contents based on  cognitive-behavioral mining in an autonomous manner without any interaction by vehicle driver during the driving.    
	\begin{remark}
		The proposed on-vehicle device application assures the recognition rather than the recall \cite{budiu2014memory} based recommendation toward the drive safe for the vehicle driver in an intelligent transportation cyber-physical system. 
	\end{remark}
\end{itemize}

A summary of the considered technologies and methodologies are illustrated in Table \ref{tab:technology_mapping}. The proposed drive safe platform meets the cutting-edge technologies and methodologies for the cognitive-behavioral mining of the vehicle driver that fulfills the goal of an intelligent transportation cyber-physical system. 
\section{Conclusion}
\label{sec:conclusion}
In this work, we have introduced a new drive safe platform that can meet the requirements of intelligent transportation cyber-physical systems towards road safety. In particular, we have proposed a cognitive-behavioral mining based driver safety platform with a working prototype that converges with the goal of the academic and industry research. Further, the driver safety platform is supported by various cutting-edge technologies, such as cognitive-behavioral mining, multi-access edge computing, artificial intelligence, and heterogeneous communications. The user study has also shown the significance of the developed drive safe prototype of IT-CPS. In the future, we will focus on improving the computational and communication latency for the vehicle driver's cognitive-behavioral mining by designing a more robust mechanism.

% or


 \appendices
 \section{Observational questionnaires for the usability test of drive safe prototype}
 \label{apx:Observational_questionnaires}
 \begin{table}[!h]
 	\caption{Observational questionnaires for the usability test of drive safe prototype }
 	\centering
 	\begin{tabular}{|p{.6cm}|p{4.5cm}|c|c|c|c|c|}
 		\hline
 		\textbf{Q. ID} &  \textbf{Observation}&\textbf{1} &\textbf{2} &\textbf{3} &\textbf{4} &\textbf{5}\\ 
 		\hline
 		H1 &  Does the on-vehicle dashboard can interact with users by always visible and visual feedback? &&&&& \\ 
 		\hline
 		H2 & Does the system can interact with vehicle driver and internal system functions, such as activity recognition, content recommender, and so on? &&&&&\\ 
 		\hline
 		H3 &  Does the system automatically provide the recommendation?  &&&&&\\ 
 		\hline
 		H4 & Does the system behavior is predictable? &&&&&\\ 
 		\hline
 		H5 &  Does the system provide audio and visual feedback to a driver?&&&&&\\ 
 		\hline
 		H6 & Does the system visualize the driver’s mood swing? &&&&&\\ 
 		\hline
 		\end{tabular}
 	\label{tab:Questionnaires_Usability_Test}
 \end{table}
1: Not good, 2: Somehow good, 3: Good, 4: Satisfactory, 5: Very satisfactory 

 \section{Questionnaires for the overall system evaluation of drive safe prototype}
  \label{apx:overall_questionnaires}
\begin{table}[!h]
	\caption{Questionnaires for the overall system evaluation of drive safe prototype}
	\centering
	\begin{tabular}{|p{.6cm}|p{6cm}|c|c|}
		\hline
		\textbf{Q. ID} &  \textbf{Question}&\textbf{0} &\textbf{1} \\ 
		\hline
		Q1 &  Would you recommend affective mood repairment on-vehicle dashboard to a friend? && \\ 
		\hline
		Q2 & Would you think affective mood repairment on-vehicle dashboard is helpful? &&\\ 
		\hline
		Q3 &  Would you think the system does not affect during the driving?  &&\\ 
		\hline
		Q4 & Does the mood swing state chart is helpful for mental safety measurement? &&\\ 
		\hline
		Q5 &  Does the autonomous audio play is a suitable feature during the driving?&&\\ 
		\hline
		Q6 & Does the system visualize the driver's mood swing? &&\\ 
		\hline
		Q7 & Does the system is useful for all kind of vehicles? &&\\ 
		\hline
		Q8 & Does the system easy to use? &&\\ 
		\hline
	\end{tabular}
	\label{tab:Questionnaires_Effectiveness_Test}
\end{table}
0: No, 1: Yes

% % use section* for acknowledgment
% \section*{Acknowledgment}


\ifCLASSOPTIONcaptionsoff
  \newpage
\fi



%\bibliographystyle{IEEEtran}


\bibliographystyle{IEEEtran}
\bibliography{mybibliography}
% \printbibliography

\begin{IEEEbiography}[{\includegraphics[width=1in,height=1.25in,clip,keepaspectratio]{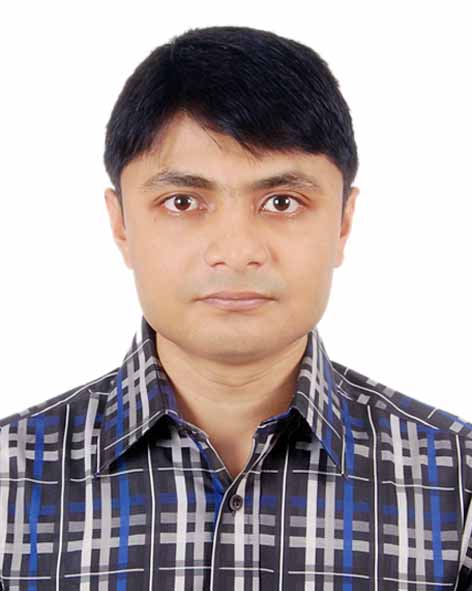}}]{Md.~Shirajum~Munir}
	(S'19) received the B.S. degree in computer science and engineering from Khulna
	University, Khulna, Bangladesh, in 2010. He is currently pursuing the Ph.D. degree in computer science and engineering at Kyung Hee University, Seoul,
	South Korea. He served as a Lead Engineer with the Solution	Laboratory, Samsung Research and Development Institute, Dhaka, Bangladesh, from 2010 to 2016. His current research interests include IoT network management, fog computing, mobile edge computing, software-defined networking, smart grid, and machine learning.
\end{IEEEbiography}

\begin{IEEEbiography}[{\includegraphics[width=1in,height=1.25in,clip,keepaspectratio]{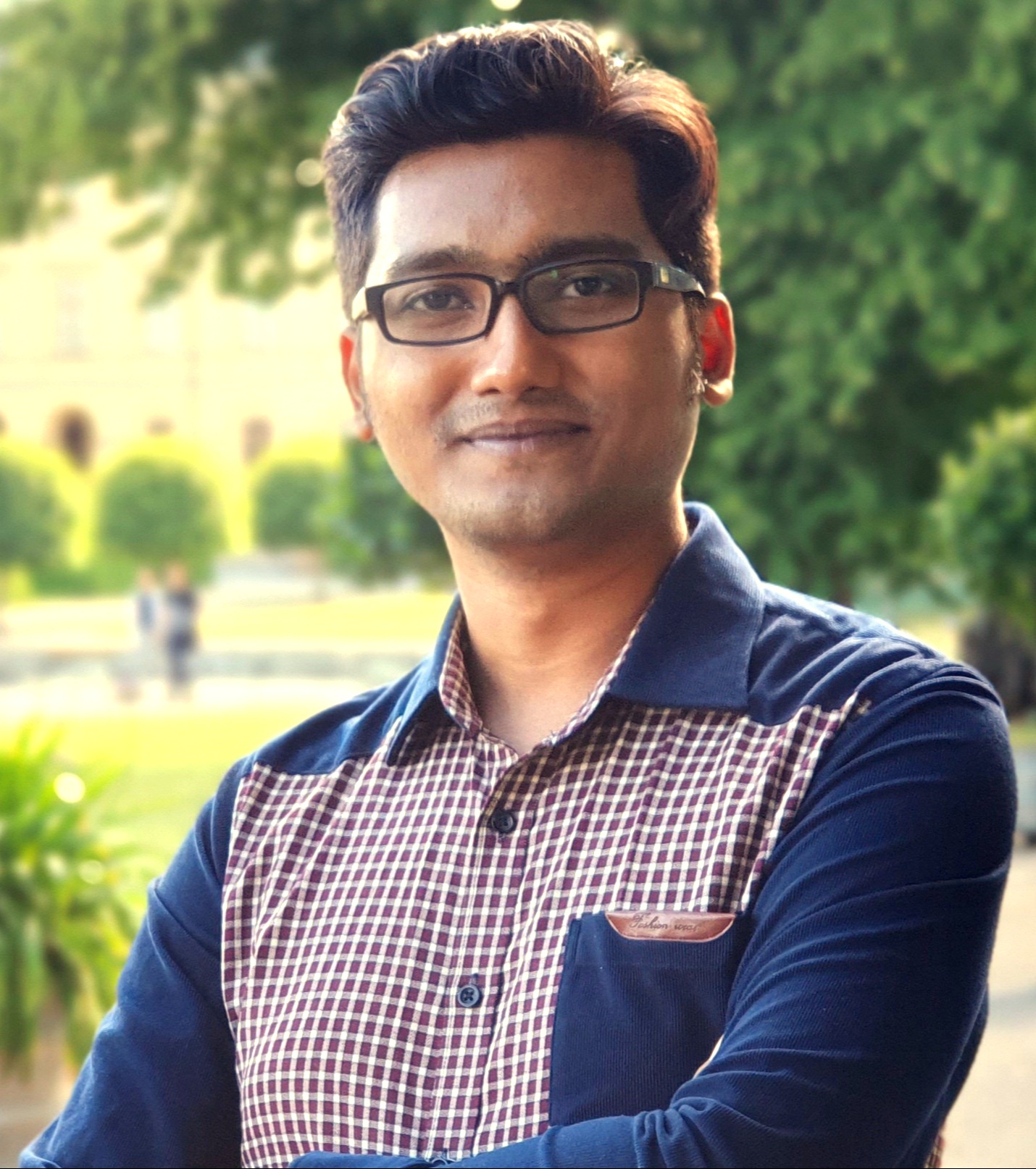}}]{Sarder~Fakhrul~Abedin}
(S'18) received his B.S. degree in Computer Science from Kristianstad University, Kristianstad, Sweden, in 2013. He received his Ph.D. degree in Computer Engineering from Kyung Hee University, South Korea in 2020. He served as a Postdoctoral Researcher at the Department of Computer Science and Engineering, Kyung Hee University, Korea. Currently, he is serving as a Postdoctoral Researcher at the
Department of Information Systems and Technology, Mid Sweden University, Sweden. His research interests include Internet of Things (IoT) network management, Edge computing, Industrial 5G, Machine learning, and Wireless networking. 
\end{IEEEbiography}

\begin{IEEEbiography}[{\includegraphics[width=1in,height=1.25in,clip,keepaspectratio]{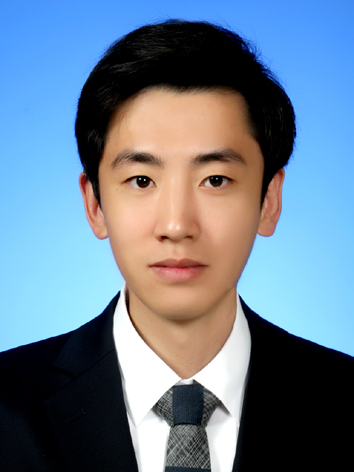}}]{Ki~Tae~Kim}
	received the B.S. and M.S. degrees in computer science and engineering from Kyung Hee University, Seoul, South Korea, in 2017 and 2019, respectively, where he is currently pursuing the
	Ph.D. degree in computer science and engineering. His research interests include SDN/NFV, wireless networks, unmanned aerial vehicle communications, and machine learning.
\end{IEEEbiography}

\begin{IEEEbiography}[{\includegraphics[width=1in,height=1.25in,clip,keepaspectratio]{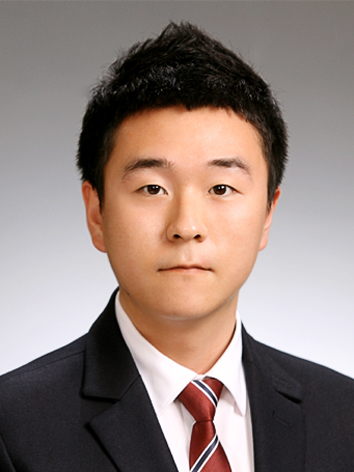}}]{Do~Hyeon~Kim}
	received the B.S. degree in communication engineering from Jeju National University, in 2014, and the M.S. degree from Kyung Hee University, in 2017, where he is currently pursuing the Ph.D. degree with the Department of Computer Science and Engineering. His research interests include multiaccess edge computing and wireless network virtualization.
\end{IEEEbiography}

\begin{IEEEbiography}[{\includegraphics[width=1in,height=1.25in,clip,keepaspectratio]{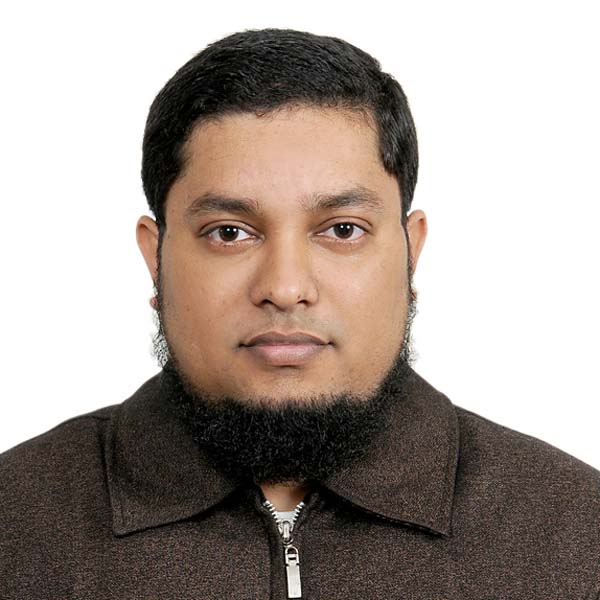}}]{Md.~Golam~Rabiul~Alam}
	(S'15-M'17) received B.S. and M.S. degrees in Computer Science and Engineering, and Information Technology respectively. He received Ph.D. in Computer Engineering from Kyung Hee University, South Korea in 2017. He served as a Post-doctoral researcher in Computer Science and Engineering Department, Kyung Hee University, Korea from March 2017 to February 2018. He is currently an Associate Professor in the Computer Science and Engineering Department at BRAC University, Bangladesh. His research interest includes healthcare informatics, mobile cloud and Edge computing, ambient intelligence, and persuasive technology. He is a member of IEEE IES, CES,
	CS, SPS, CIS, and ComSoc. He is also a member of the Korean Institute of Information Scientists and Engineers (KIISE) and received several best paper awards from prestigious conferences.
\end{IEEEbiography}

\begin{IEEEbiography}[{\includegraphics[width=1in,height=1.25in,clip,keepaspectratio]{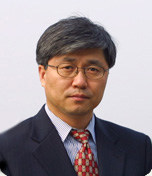}}]{Choong~Seon~Hong}
	(S'95-M'97-SM'11)
	received the B.S. and M.S. degrees in electronic engineering from Kyung Hee University, Seoul, South Korea, in 1983 and 1985, respectively, and the Ph.D. degree from Keio University, Japan, in 1997. In 1988, he joined KT, where he was involved in broadband networks as a Member of Technical Staff. Since 1993, he has been with Keio University. He was with the Telecommunications Network Laboratory, KT, as a Senior Member of Technical Staff and as the Director of the Networking Research Team until 1999. Since 1999, he has been a Professor with the Department of Computer Science and Engineering, Kyung Hee University. His research interests include future Internet, ad hoc networks, network management, and network security. He is a member of the ACM, the IEICE, the IPSJ, the KIISE, the KICS, the KIPS, and the OSIA. Dr. Hong has served as the General Chair, the TPC Chair/Member, or an Organizing Committee Member of international conferences such as NOMS, IM, APNOMS, E2EMON, CCNC, ADSN, ICPP, DIM, WISA, BcN, TINA, SAINT, and ICOIN. He was an Associate Editor of the IEEE TRANSACTIONS ON NETWORK AND SERVICE MANAGEMENT, and the IEEE JOURNAL OF COMMUNICATIONS AND NETWORKS. He currently serves as an Associate Editor of the International Journal of Network Management, and an Associate Technical Editor of the IEEE Communications Magazine.
\end{IEEEbiography}

% \end{IEEEbiography}

% bio text here

%\newpage



%\vfill


% \bibliographystyle{IEEEtran}
% \printbibliography
\end{document}